\documentclass[a4paper,fleqn,usenatbib]{mnras}

\usepackage{newtxtext,newtxmath}

\usepackage[T1]{fontenc}
\usepackage{ae,aecompl}

\usepackage{graphicx}	
\usepackage{amsmath}	
\usepackage{amssymb}	
\usepackage{pdflscape}

\usepackage{graphicx}
\usepackage{epsfig}
\usepackage{multirow}
\usepackage{float}
\usepackage{amsmath}
\usepackage{arydshln}
\usepackage{afterpage}
\def\mnras{MNRAS}
\def\apj{ApJ}
\def\aj{AJ}
\def\apjl{ApJL}
\def\apjs{ApJS}
\def\aap{A\& A}

\def\pasj{PASJ}
\def\nat{Nature}
\def\axj{AX J1745.6-2901}
\def\sgra{Sgr A$^\star$}
\def\xmm{{\it XMM-Newton}}
\def\cxo{{\it Chandra}}
\def\swift{{\it Swift}}
\def\rosat{{\it ROSAT}}

\title[Effects of Dust Scattering on the X-ray Eclipses]{Effects of Interstellar Dust Scattering on the X-ray Eclipses of the LMXB \axj\ in the Galactic Center}

\author[C. Jin et al.]{
Chichuan Jin$^{1,2}$\thanks{E-mail: chichuan@mpe.mpg.de},
Gabriele Ponti$^{1}$,
Frank Haberl$^{1}$,
Randall Smith$^{3}$,
Lynne Valencic$^{4}$
\\
$^{1}$Max-Planck-Institut f\"{u}r Extraterrestrische Physik, Giessenbachstrasse, D-85748 Garching, Germany\\
$^{2}$National Astronomical Observatories, Chinese Academy of Sciences, A20 Datun Road, Beijing 100101, China\\
$^{3}$Smithsonian Astrophysical Observatory, 60 Garden Street, Cambridge, MA 02138, USA\\
$^{4}$Johns Hopkins University, 3400 N. Charles St., Baltimore, MD 21218, USA\\
}

\date{prepared for MNRAS}

\pubyear{2018}

\begin{document}
\label{firstpage}
\pagerange{\pageref{firstpage}--\pageref{lastpage}}
\maketitle

\begin{abstract}
\axj\ is an eclipsing low mass X-ray binary (LMXB) in the Galactic Centre (GC). It shows significant X-ray excess emission during the eclipse phase, and its eclipse light curve shows an asymmetric shape. We use archival \xmm\ and \cxo\ observations to study the origin of these peculiar X-ray eclipsing phenomena. We find that the shape of the observed X-ray eclipse light curves depends on both photon energy and the shape of the source extraction region, and also shows differences between the two instruments. By performing detailed simulations for the time-dependent X-ray dust scattering halo, as well as directly modelling the observed eclipse and non-eclipse halo profiles of \axj, we obtained solid evidence that its peculiar eclipse phenomena are indeed caused by the X-ray dust scattering in multiple foreground dust layers along the line-of-sight (LOS). The apparent dependence on the instruments is caused by different instrumental point-spread-functions. Our results can be used to assess the influence of dust scattering in other eclipsing X-ray sources, and raise the importance of considering the timing effects of dust scattering halo when studying the variability of other X-ray sources in the GC, such as \sgra. Moreover, our study of halo eclipse reinforces the existence of a dust layer local to \axj\ as reported by Jin et al. (2017), as well as identifying another dust layer within a few hundred parsecs to Earth, containing up to several tens of percent LOS dust, which is likely to be associated with the molecular clouds in the Solar neighbourhood. The remaining LOS dust is likely to be associated with the molecular clouds located in the Galactic disk in-between.
\end{abstract}

\begin{keywords}
dust scattering, Galactic Center, eclipsing X-ray binaries: AX J1745.6-2901
\end{keywords}

\section{Introduction}
\label{sec-intro}
\subsection{X-ray Light Echoes from the Dust Scattering}
\label{sec-intro-lightecho}
\begin{figure}
\includegraphics[bb=-10 0 612 480,scale=0.242]{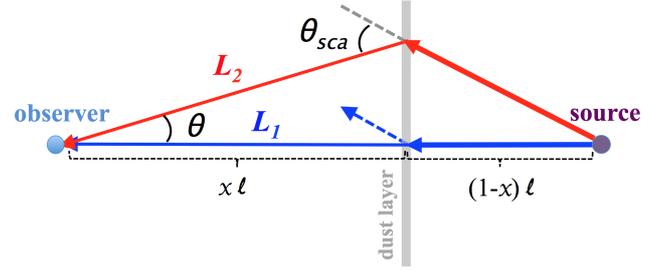}
\caption{Geometry of the dust scattering by a single dust layer. $\theta$ is the viewing angle. $\theta_{\rm sca}$ is the scattering angle. $\ell$ is the absolute distance between the source and the observer. $x$ is the fractional distance from the dust layer to the observer. $L_1$ and $L_2$ signify the two different light paths traveled by a LOS photon and a scattered photon in the halo.}
\label{fig-cartoon1}
\end{figure}

X-ray dust scattering is an astrophysical phenomenon known for decades (Overbeck 1965; Tr\"{u}mper \& Sch\"{o}nfelder 1973). It is a typical small-angle elastic scattering well described by the Mie scattering theory. The observational effects of X-ray dust scattering by the interstellar medium (ISM) comprise several aspects. Firstly, the foreground dust grains of an X-ray source will scatter some of the line-of-sight (LOS) photons into other directions, thereby dimming the source. Secondly, dust grains outside the source LOS can scatter X-ray photons in the same way, thereby redirecting some source photons to the observer on Earth, as shown in Fig.\ref{fig-cartoon1}. If the point-like X-ray source is persistent, these photons will create an extended halo around it, i.e. the so-called X-ray dust scattering halo. The intensity and radial profile of the halo can provide important clues about the dust grains in the ISM, such as the grain size distribution, spatial distribution and optical depth ($\tau_{\rm sca}$, e.g. Mauche \& Gorenstein 1986; Mathis \& Lee 1991; Smith \& Dwek 1998; Draine 2003). Dust scattering can also change the observed source spectrum, because the scattering opacity has an energy dependence, and it is always compensated by the partial inclusion of halo photons (e.g. Predehl \& Schmitt 1995; Smith, Valencic \& Corrales 2016; Corrales et al. 2016; Jin et al. 2017, hereafter: J17).

The first observation of an X-ray dust scattering halo was made by the {\it Einstein} satellite from the Galactic low-mass X-ray binary (LMXB) GX 339-4 (Rolf 1983). Then X-ray dust scattering halos were observed around many other sources by the following X-ray satellites with higher spatial resolution and larger effective area, such as \rosat, \swift, \cxo\ and \xmm\ (e.g. Predehl \& Schmitt 1995; Xiang, Zhang \& Yao 2005; Valencic \& Smith 2015). Together with the measurement of optical extinction (e.g. the V-band extinction: $A_V$) and X-ray absorption (e.g. the equivalent hydrogen column: $N_{\rm H,abs}$), dust-to-gas relations in the ISM have been well-established between $\tau_{\rm sca}$, $A_V$ and $N_{\rm H,abs}$, which have wide astronomical applications (Predehl \& Schmitt 1995; Valencic \& Smith 2015).

If the primary X-ray source is variable, the dust scattering halo will also vary, but with a time lag, which is due to the different light paths travelled by the LOS photons and the halo photons. For example, the two light paths of $L_1$ and $L_2$ in Fig.\ref{fig-cartoon1} differ by
\begin{equation}
\label{equ-1}
\Delta{\ell}(x,\theta)\equiv L_2-L_1=\Big{\{}\sqrt{\Big{(}\frac{x}{cos\theta}\Big{)}^2-2x+1}+\frac{x}{cos\theta}-1\Big{\}}\cdot\ell
\end{equation}
where $\theta$ is the viewing angle and $x$ is the fractional distance to the observer (so the source is at $x=1$). After applying small-angle approximations, the corresponding time lag can be expressed as
\begin{equation}
\label{equ-2}
\Delta t(x,\theta,\ell)\equiv\frac{\Delta\ell}{c}\approx1.21 ({\rm s})~\frac{x}{1-x}~\theta({\rm arcsec})^2~\ell({\rm kpc})
\end{equation}
where $c$ is the speed of light (Xu, McCray \& Kelley 1986). Thus for an X-ray source located 10 kpc away with a single foreground geometrically thin dust layer at $x=0.5$, scattering photons from a viewing angle of 1-60 arcsec would lag behind the LOS photons by 12.1 s - 43.6 ks.

Therefore, if an X-ray source produces an energetic short outburst (i.e. a pulsed signal), each of the foreground thin dust layers will produce an expanding dust scattering ring for this outburst, thus a set of expanding rings will emerge around the source after the outburst. Indeed, such dust scattering ring structures have been observed around several X-ray sources, such as V404 Cygni (Vasilopoulos \& Petropoulou 2016; Beardmore et al. 2016; Heinz et al. 2016), Circinus X-1 (Heinz et al. 2015), IGR J17544-2619 (Mao, Ling \& Zhang 2014), 1E 1547.0-5408 (Tiengo et al. 2010) and SGR 1806-20 (Svirski et al. 2011). These expanding rings can be used to measure the absolute distance to the source and multiple foreground dust layers.

However, if the X-ray source is varying continuously rather than producing pulsed signals, it would be very difficult to observe expanding ring structures directly. Instead, a dust scattering halo with variable radial profile is expected. Assuming that there is no azimuthal variation of the foreground dust distribution, the time-dependent halo profile can be expressed as
\begin{multline}
\label{equ-3}
I_{\rm sca}(\theta, E, \ell, t)=N_{\rm H,sca} {\int^1_0} F(t-\Delta t)~\cdot {\frac{f(x)}{(1-x)^2}} \\
\times {\int^{E_{\rm max}}_{E_{\rm min}}} S(E) {\int^{a_{\rm max}}_{a_{\rm min}}} n(a)~{\frac{d\sigma_{\rm sca}(a,x,\theta,E)}{d\Omega}}~da~dx~dE
\end{multline}
where $I_{\rm sca}$ is the halo intensity due to single dust scatterings, which normally dominate higher-order scatterings for X-ray photons above 2 keV (Mathis \& Lee 1991; Xiang, Lee \& Nowak 2007; J17). $a$ is the size of a dust grain, $E$ is the photon energy. $N_{\rm H, sca}$ is the equivalent hydrogen column for the dust scattering opacity, which depends on the abundances and dust-to-gas ratio defined in a dust grain model. $F(t)$ is the intrinsic source light curve. $\Delta t$ is the time lag in Eq.\ref{equ-2}. $f(x)$ is the normalized spatial dust distribution along the LOS. $S(E)$ is the normalized source spectrum, which can be approximated by a spectra-weighted effective photon energy for the halo calculation (see J17). $d\sigma_{\rm sca}/d{\Omega}$ is the differential scattering cross-section, whose expression can be found in J17 (Equation 2-4). $n(a)$ is the dust grain size distribution, which depends on the adopted dust grain model. Note that many dust grain models have been developed so far, each with different dust grain compositions, size distributions, abundances and dust-to-gas ratios (e.g. Mathis, Rumpl \& Nordsieck 1977; Weingartner \& Draine 2001; Zubko, Dwek \& Arendt 2004; Xiang et al. 2011), and so $n(a)$ and $N_{\rm H, sca}$ often contain large systematic uncertainties (see J17 and references therein).

\subsection{X-ray Dust Scattering towards the Galactic Centre}
The Galactic Centre region (hereafter: GC) is heavily extinguished from the optical to X-ray band (e.g. Becklin \& Neugebauer 1968; Fritz et al. 2011). Assuming common dust-to-gas relations (e.g. G\"{u}ver \& \"{O}zel 2009; Zhu et al. 2017), we expect a large amount of intervening dust along the GC LOS, and these dust grains should introduce dust scattering effects to all X-ray sources in the GC. However, there were very few works about the dust scattering towards the GC, including the early work by Tan \& Draine (2004) on \sgra, the work by Corrales et al. (2017) who reported the existence of a dust scattering halo around the GC transient SWIFT J174540.7-290015 (also see Ponti et al. 2016), and the work by J17 where the first detailed analysis of the dust scattering halo was performed for a bright X-ray source in the GC LOS, namely \axj\ (Maeda et al. 1996).

\subsubsection{Dust Scattering towards \sgra}
Tan \& Draine (2004) first calculated the response of foreground dust scattering to the flare of \sgra, where they considered most of the dust to be relatively close to \sgra. However, recent studies about the interstellar gas scattering towards the magnetar SGR J1745-2900, which is located 2.4 arcsec away from \sgra, favour the scattering source to be far from the GC (Rea et al. 2013; Eatough et al. 2013; Shannon \& Johnston 2013; Bower et al. 2014; Wucknitz 2015; Sicheneder \& Dexter 2017). Besides, massive molecular clouds have also been found to reside along the spiral arms between Earth and the GC (Simon et al. 2006; Marshall, Joncas \& Jones 2009; Roman-Duval et al. 2010; Dobbs \& Burkert 2012; Sato et al. 2014; Heyer \& Dame 2015). From the dust scattering halo study of \axj, J17 reported that most of the GC foreground dust may reside in the Galactic disk. Since the angular separation between \axj\ and \sgra\ are so small, and their $N_{\rm H, sca}$ values are also similarly high (Ponti et al. 2017b; J17), their LOS dust distribution is likely to be very similar to each other, so it is indeed likely that most of the foreground dust towards \sgra\ is in the Galactic disk rather than in the GC (J17).

\begin{table*}
 \centering
  \begin{minipage}{175mm}
  \centering
   \caption{List of eclipse periods of \axj\ detected by \xmm\ and \cxo. We only chose observations in J17 where \axj\ was bright enough to allow the study of the eclipse light curve (see Section~\ref{sec-data}). The list of observations used to determine the underlying diffuse emission and detector background can be found in Table B1, B2 in J17. $\theta_{\rm off}$ is the off-axis angle of \axj\ relative to the pointing direction of the optical axis. PFW-Medium: `Prime-Full-Window' mode with the `Medium' filter. HETG: High Energy Transmission Grating.}
    \label{tab-obs}
     \begin{tabular}{@{}ccccccc@{}}
     \hline
     Satellite & ObsID  & Instrument & Obs-Mode & $\theta_{\rm off}$ & Eclipse Ingress Time & Eclipse Egress Time\\
     & & & & (arcmin) & (MJD) & (MJD)\\
     \hline
     \xmm\ & 0402430701 & EPIC-pn &PFW-Medium&2.57 & 54190.05919& 54190.07556\\
     \xmm\ & 0402430301 & EPIC-pn &PFW-Medium&2.55 & 54191.79898& 54191.81542\\
     \xmm\ & 0402430301 & EPIC-pn &PFW-Medium&2.55 & 54192.49482& 54192.51129\\
     \xmm\ & 0402430401 & EPIC-pn &PFW-Medium&2.60 & 54193.88681& 54193.90319\\
     \xmm\ & 0402430401 & EPIC-pn &PFW-Medium&2.60 & 54194.23470& 54194.25111\\
     \xmm\ & 0402430401 & EPIC-pn &PFW-Medium&2.60 & 54194.58267& 54194.59912\\
     \xmm\ & 0505670101 & EPIC-pn &PFW-Medium&2.57 & 54194.25111& 54549.16899\\
     \xmm\ & 0724210201 & EPIC-pn &PFW-Medium&1.94 & 54549.50045& 54549.51691\\
     \xmm\ & 0700980101 & EPIC-pn &PFW-Medium&1.92 & 56545.39197& 56545.40833\\
     \xmm\ & 0724210501 & EPIC-pn &PFW-Medium&1.89 & 56558.26641& 56558.28280\\
     \xmm\ & 0743630801 & EPIC-pn &PFW-Medium&2.56 & 57113.60843& 57113.62483\\
     \hline
     \cxo\ & 17857 & ACIS-S & HETG&0.34& 57245.84512 & 57245.86225\\
     \cxo\ & 17857 & ACIS-S & HETG&0.34& 57246.19396 & 57246.21086\\
     \cxo\ & 17857 & ACIS-S & HETG&0.34& 57246.54211 & 57246.55866\\
     \cxo\ & 17857 & ACIS-S & HETG&0.34& 57246.88956 & 57246.90658\\
     \hline
     \end{tabular}
 \end{minipage}
\end{table*}

\subsubsection{Dust Scattering towards \axj}
\label{sec-intro-axj1745}
\axj\ is one of the brightest X-ray transients located 87.44 arcsec away from \sgra\ (Porquet et al. 2008; Petrov et al. 2011; Ponti et al. 2015; J17). Taking the advantages of both \cxo's high spatial resolution and \xmm's large effective area, J17 measured the dust scattering halo around \axj\ in three different energy bands with high special resolution and signal-to-noise, and performed detailed radial profile analysis using 19 different dust grain populations in the literatures (Mathis, Rumpl \& Nordsieck 1997, hereafter: MRN77; Weingartner \& Draine 2001; Zubko, Dwek \& Arendt 2004, hereafter: ZDA04; Xiang et al. 2011). Two major geometrically thick dust layers were reported in front of \axj. Layer-1 was found to lie within a fractional distance of 0.11 to \axj\ and contain (19-34)\% of the LOS intervening dust\footnote{\label{fn-dust-frac}Two other important assumptions related to this LOS dust fraction are that J17 adopted a single dust grain population for all dust layers, and the additional dust grain population responsible for the halo wing was not taken into account.}, probably associated with the central molecular zone (hereafter: CMZ, Serabyn \& Guesten 1987; Morris \& Serabyn 1996; Ponti et al. 2013; Henshaw et al. 2016). Layer-2 is distributed from Earth up to a fractional distance of 0.64 and contains (66-81)\% of the LOS dust, and so these dust grains are likely to reside in the Galactic disk several kpc away from the GC. However, the halo profile fitting alone was not able to further resolve the dust distribution within this thick layer, thus it would be useful to study the timing properties of the halo in order to obtain further constraints on the foreground dust properties.

A special characteristic of \axj\ is that it is an eclipsing source (Maeda et al. 1996). It has an orbital period of $30063.74\pm0.14$ s and an eclipse period of $1440$ s (Hyodo et al. 2009). Recently, Ponti et al. (2017a) found a long-term decreasing rate of $(4.03\pm0.32)\times10^{-12}$ s/s in its orbital period with significant jitter. What makes \axj\ peculiar is that it shows significant excess flux during the eclipse phase, and the shape of the eclipse light curve is not symmetric (Maeda et al. 1996; Ponti et al. 2017a). In comparison, in most of other eclipsing X-ray binaries, the eclipse light curve usually drops to zero-flux with a symmetric shape (e.g. MXB 1659-29: Wachter, Smale \& Bailyn 2000; Her X-1: Leahy \& Scott 2001; Leahy 2015; GRS J1747-312: in't Zand et al. 2003; EP Dra: Ramsay et al. 2004; EXO 0748-676: Homan, Wijnands \& van den Berg 2003; Wolff et al. 2009; [PMH2004] 47: Pietsch et al. 2009; XTE J1710-281: Jain \& Paul 2011; IGRJ17451-3022: Bozzo et al. 2016). The origin of the peculiar eclipse light curve of \axj\ was not clear. Maeda et al. (1996) found that the ratio between the eclipse spectrum and the non-eclipse spectrum has an energy-dependence of $E^{-2} \sim E^{-3}$, which is roughly consistent with the energy-dependence of the dust scattering cross-section (e.g. Predehl \& Schmitt 1995), so they inferred a dust scattering origin for this excess flux in the eclipse. Another similar source is Swift J1749.4-2807, where some residual flux is also observed during its eclipse phase and a dust scattering origin was also speculated (Ferrigno et al. 2011). However, so far there is no direct evidence to verify the dust scattering origin, and the excess flux and the asymmetric shape of the eclipse light curve have never been modelled directly.

Since the dust scattering halo of \axj\ has been modelled in J17, it is now possible to study the timing properties of its foreground dust scattering and test the effect of dust scattering on the eclipse light curve. We note that \axj\ also shows complex dipping phenomena especially before the eclipse ingress time and some short type-I bursts, but the eclipse signal is the only periodic signal in its light curve, so in this work we focus on the eclipsing phenomenon. To perform modelling and simulations, we can approximate the intrinsic eclipse light curve of \axj\ with the following rectangular function
\begin{equation}
\label{equ-4}
  F(t)=\begin{cases}
    ~0 & \text{$0~{\rm s} \le t~{\rm mod}~T_{\rm orbit} < T_{\rm eclips}$}\\
    ~F_0 & \text{otherwise}
  \end{cases}
\end{equation}
where $F_{\rm 0}$ is the flux right before the eclipse ingress. $T_{\rm orbit}=30.064$ ks and $T_{\rm eclips}=1440$ s. Then Eq.\ref{equ-1}-\ref{equ-4} allow us to calculate the response of the dust scattering halo to the periodic eclipse signal.

In this paper, we study the variability of the X-ray dust scattering halo due to the eclipse signal, and directly model the observed time-dependent halo profile of \axj\ and its eclipse light curves using the dust scattering in multiple foreground dust layers. This paper is organised as follows. In Section 2 we describe the \xmm\ and \cxo\ observations used in this work, as well as the principal data reduction procedures. Section 3 presents the dependence of the eclipse light curve of \axj\ on the source extraction region (i.e. radial distance) and photon energy as observed by different instruments with different point-spread function (PSF). In Section 4 we perform simulations to understand the response of dust scattering to the eclipse signal in different dust layers, and show their characteristic eclipse light curves. Then in Section 5 we directly model the eclipse halo profile and light curves of \axj\ using dust scattering. In Section 6 we discuss further constraints obtained from our halo timing study on the GC foreground dust distribution, and discuss their association with the MCs in the Galactic disk. Summaries and conclusions are presented in Section 7. Similar to J17, we adopt a distance of 8 kpc to the GC (Reid et al. 2009; Genzel et al. 2010; Boehle et al. 2016).

\begin{figure*}
\begin{tabular}{cc}
\includegraphics[bb=90 216 576 648,scale=0.39]{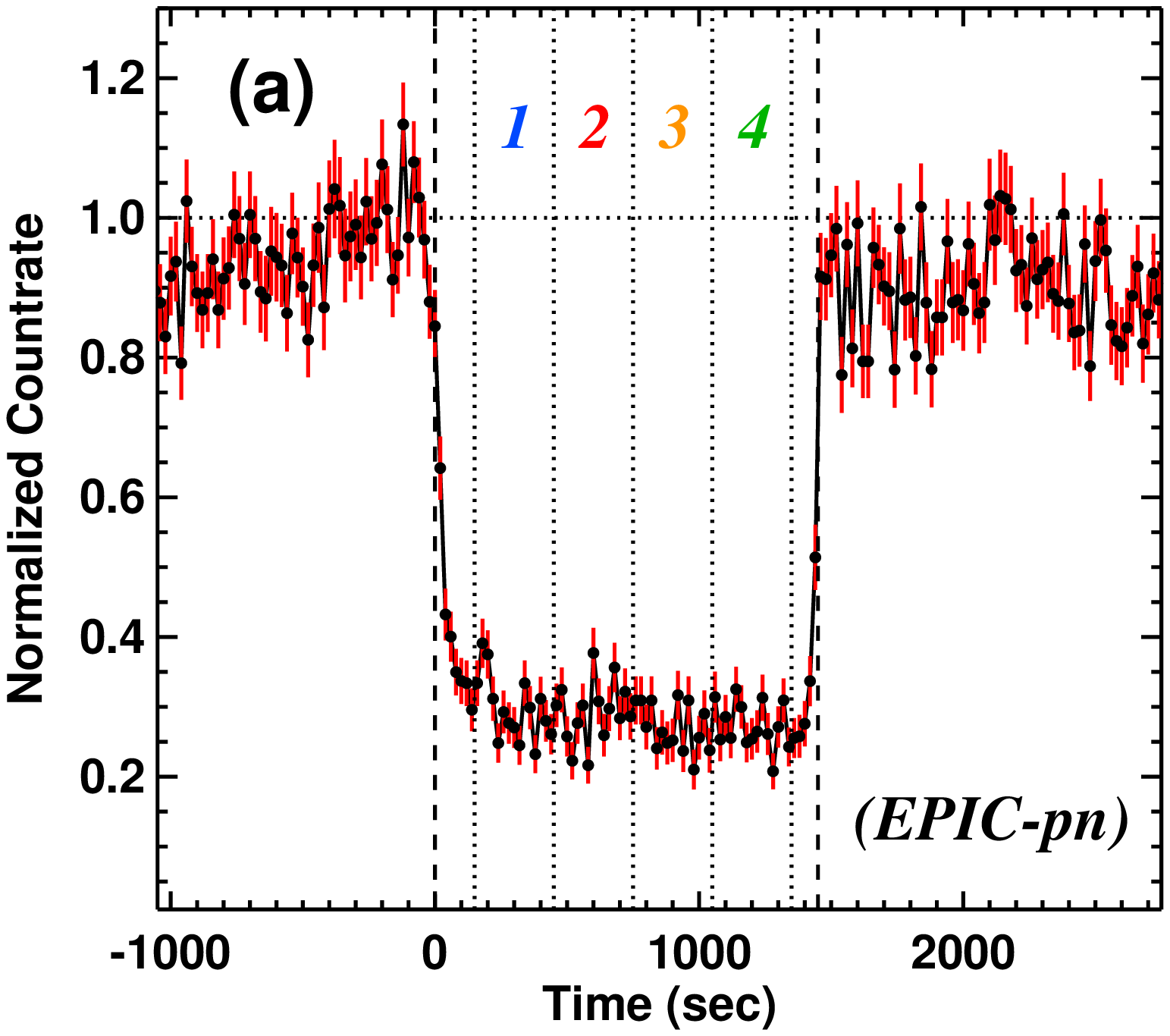} &
\includegraphics[bb=90 216 576 648,scale=0.39]{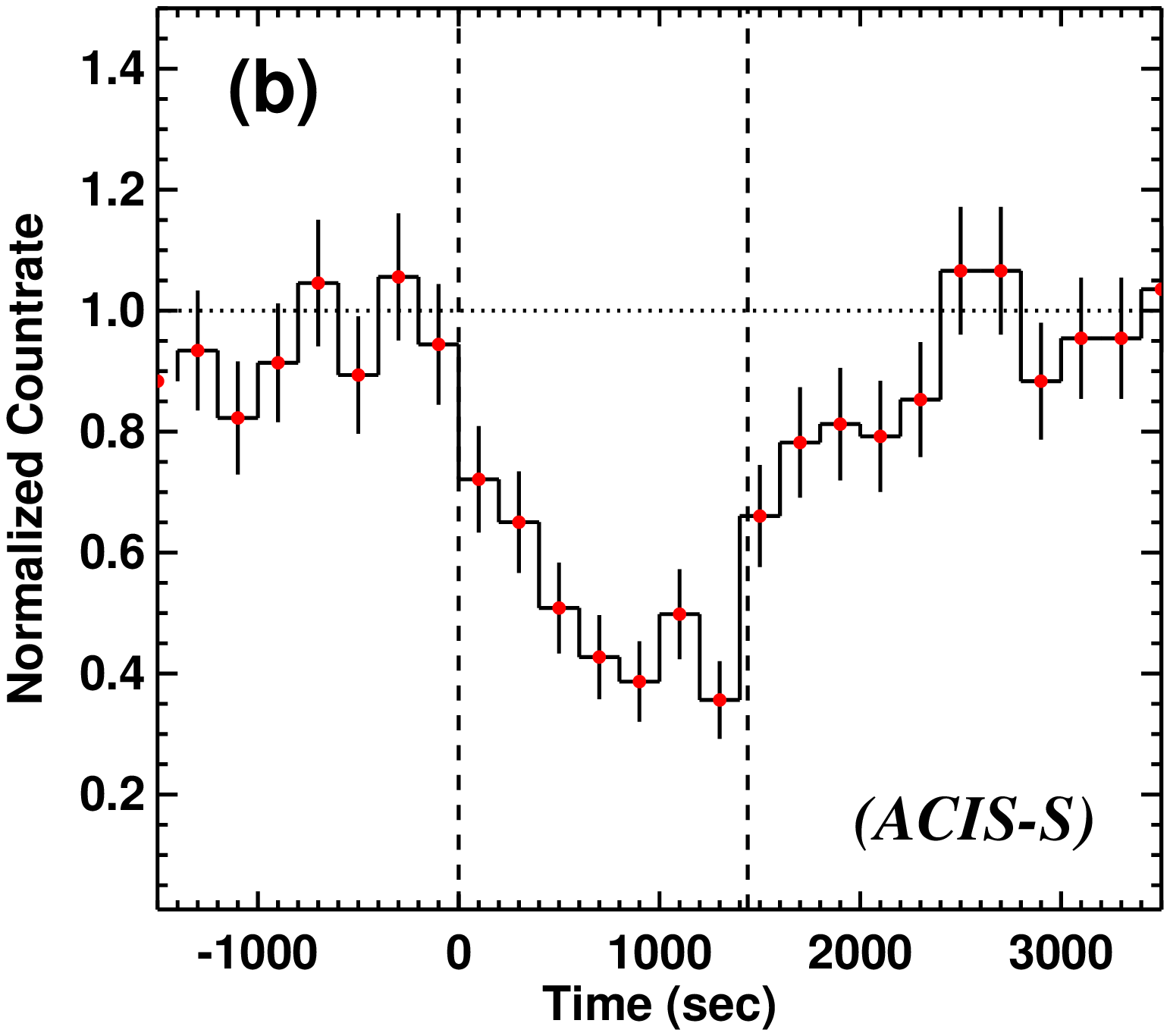} \\
\end{tabular}
\caption{The combined 2-10 keV eclipse light curves of \axj\ based on the observations listed in Table~\ref{tab-obs}. Panel-a shows the results for \xmm\ EPIC-pn, where the source extraction region is 20-100 arcsec and the light curve binning time is 20 s. The eclipse phase is divided into 4 time intervals, where the corresponding radial profiles can be found in Fig.\ref{fig-eclips-rad}. Panel-b shows the results for \cxo\ ACIS-S, where the source extraction region is 2.4-20 arcsec and the binning time is 100 s. Background count rates have been subtracted, which were determined from observations during which \axj\ was in quiescence (see Table B1, B2 in J17). $t=$ 0 s is the eclipse ingress time.}
\label{fig-lc210}
\end{figure*}

\begin{figure*}
\begin{tabular}{c}
\includegraphics[bb=200 0 500 792, scale=0.5, angle=90]{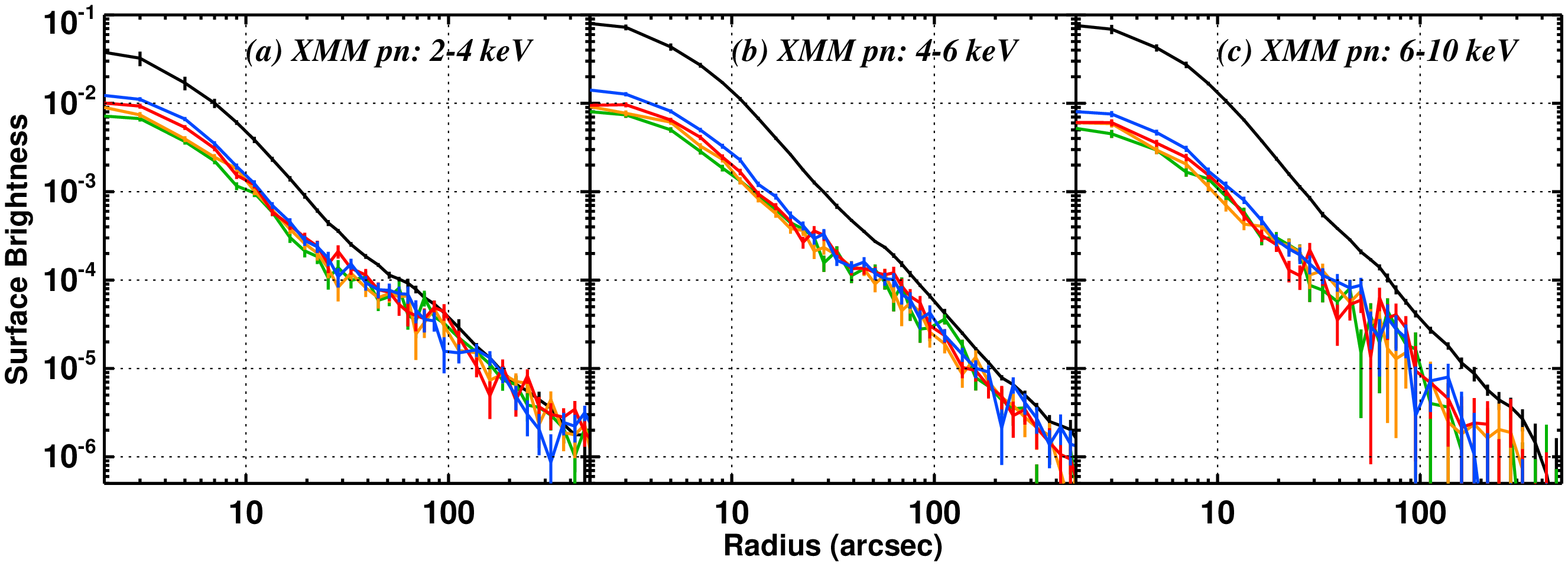}\\
\includegraphics[bb=210 0 470 792, scale=0.5, angle=90]{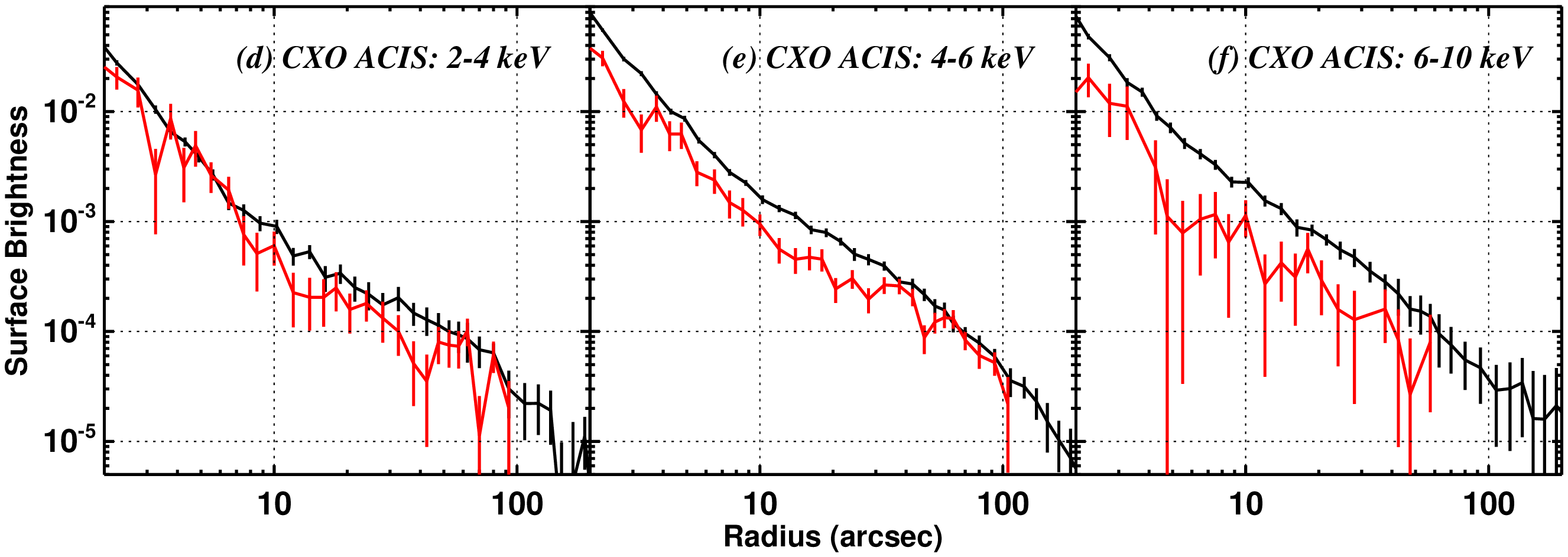}\\
\end{tabular}
\caption{Comparison of the radial profiles of \axj\ during the non-eclipse and eclipse phases for the energy bands of 2-4, 4-6 and 6-10 keV. In Panel-a,b,c the black line is the radial profile during the non-eclipse phase as being observed by \xmm\ EPIC-pn. The blue, red, orange and green profiles are the radial profiles observed during the four time intervals within the eclipse phase in Fig.\ref{fig-lc210}a. The central 20 arcsec was contaminated by photon pile-up; we only plot them for the purpose of comparison. In Panel-d,e,f the black line is the radial profile observed during the non-eclipse phase by \cxo\ ACIS-I, and renormalized to the brightness of those observed during the ACIS-S observation. We did not directly use the non-eclipse radial profiles from ACIS-S because of the smaller radial coverage in the sub-array mode, and they are consistent with these ACIS-I radial profiles (see J17). The red line is the eclipse radial profiles observed by \cxo\ ACIS-S. Background radial profiles were determined from observations during which \axj\ was in quiescence (see Table B1, B2 in J17), and were subtracted from these radial profiles.}
\label{fig-eclips-rad}
\end{figure*}

\section{Observation and Data Reduction}
\label{sec-data}
The eclipse of \axj\ has been observed many times by \xmm\ and \cxo\ during the past twenty years (see Ponti et al. 2017a and references therein). Since this work mainly focuses on the eclipsing phenomena, a high source count rate is required to increase the signal-to-noise during the eclipse phase, so we chose a subset of observations in J17 during which \axj\ was observed at a high flux level (i.e. $F_{\rm 2-4~keV} \ge 10^{-11}$ erg cm$^{-2}$ s$^{-1}$). Among them we chose observations covering at least one complete eclipse phase, which resulted in 8 \xmm\ observations covering 11 complete eclipse periods and one \cxo\ observation covering 4 eclipse periods (see Table~\ref{tab-obs}). The ingress and egress time of these eclipse periods have been reported by Ponti et al. (2017a). A large set of archival \xmm\ and \cxo\ observations during which \axj\ was in quiescence was used to accurately determine the strong GC diffuse emission and detector background underneath \axj. In this section we only summarise the principal data reduction procedures and methods, while a full description can be found in J17.

\subsection{\xmm}
For \xmm\ observations (Jansen et al. 2001), the Science Analysis System ({\sc sas} v16.0.0) was used to reduce the data. We mainly focused on the data from EPIC-pn because it provides the highest count rate among all three EPIC cameras. The {\sc epchain} task was used to reprocess the data with the latest calibration files. The data periods contaminated by strong background flares were excluded (e.g. Urban et al. 2011). The Extended Source Analysis Software ({\sc esas}, Snowden \& Kuntz 2011) was used to create images (including source images, particle background images, Out-of-Time (OoT) processing images) and exposure maps (including the vignetting effect) in different energy bands, and was also used to perform the point source detection. Then the OoT images and particle background images were subtracted from the source images using the {\sc ftools} (v6.19) package. J17 used the {\sc epatplot} task and the method of Costantini, Freyberg \& Predehl (2005) to check the photon pile-up, and found that pile-up only affected the region within 20 arcsec of \axj. Thus although the data was also extracted from this circular region, we did not model them directly in this paper. Spectra were extracted using the {\sc sas} {\sc evselect} task, and then the {\sc rmfgen} and {\sc arfgen} tasks were used to create the response files and unvignetted ancillary files which contain the spectral weighted effective areas.

We followed the same approach as being described in J17 to derive the radial profile of \axj. Firstly, we masked out regions containing point sources, strong diffuse emission, weak stray lights and artificial features such as the residual OoT feature. Then the radial profile of \axj\ was extracted within 600 arcsec from every observation using the {\sc DS9} application. To get the combined radial profile, in every annulus we added the counts from all the observations, and divided the total counts by the total exposure time and the effective area to get the surface brightness. The error bar for the combined radial profile includes both the statistical error and the small dispersion between different observations, which is likely due to the small variation of PSF profile for slightly different off-axis angles\footnote{We notice that the off-axis angles in Table 1 of J17 were not accurate. Now the better measured values are reported in Table~\ref{tab-obs} of this paper.} and the residual background contaminations. This procedure was repeated for both source and background observations, then the combined background radial profile was subtracted from the combined source radial profile to derive the intrinsic radial profile of \axj\ (see J17 for a full description).

Our study also made use of light curves from multiple observations of \axj, so it is necessary to perform the barycenter correction on the event file of every observation using the {\sc sas} {\sc barycen} task. This eliminated the difference of the photon arrival time between different observations due to the orbital motion of Earth and satellite around the Sun. Then light curves were extracted from the event file using the {\sc evselect} task for different energy bands and annular regions around the source. Similar to the radial profile construction, the background flux was measured from observations during which \axj\ was in quiescence and subtracted from the source light curves. Based on the ingress time of each eclipse period, light curves from different observations were combined to produce the folded eclipse light curves with Poisson errors. The PSF profiles of EPIC cameras reported by Ghizzardi (2002), which were based on 110 \xmm\ observations and have both energy and off-axis angle, were used in this work to perform halo simulations and modelling. Since all the \xmm\ observations used in this work have very small off-axis angles (see Table~\ref{tab-obs}), the effect of PSF's dependence on the off-axis angle is not important compared to the statistical errors, and it is also taken into account in the error bar of the combined radial profiles. Moreover, Ghizzardi (2002) showed the consistency between the data and their best-fit PSF model up to $\sim$300 arcsec, while in this paper we only modelled the central 200 arcsec of the halo, so the uncertainty of calibration in the extended PSF wing should not affect our results.

\subsection{\cxo}
For \cxo\ observations, the {\sc ciao} (\cxo\ Interactive Analysis of Observations) software (v4.9) was used to reduce the data. The {\sc chandra\_repro} script was used to reprocess the ACIS data with the calibration database (CALDB v4.7.3). The {\sc axbary} script was used to perform the barycenter correction on both event files and aspect solution files{\footnote{http://cxc.harvard.edu/ciao/threads/axbary}. The {\sc merge\_obs} and {\sc reproject\_image\_grid} scripts were used to create images and exposure maps (including the vignetting effect) for every observation, and create combined images from all the observations. The {\sc pileup\_map} script was used to create pile-up maps, which were used to estimate the pile-up fraction in every pixel. As being reported by J17, the central 2.4 arcsec of \axj\ contains more than 1\% pile-up, so we excluded the data within this region. Then the {\sc wavedetect} script was used to perform the point source detection\footnote{\cxo\ resolved some faint point sources near \axj, which were not detected by \xmm\ due to the broader PSF of EPIC-pn and the presence of the dust scattering halo.}. After masking out point sources and artificial features (e.g. readout strikes and grating arms), the construction of individual and combined radial profiles were performed in the same way as for \xmm. Then the {\sc dmextract} script was used to extract light curves from every observation, which were then used to create the folded eclipse light curves. The ACIS PSF radial profile used in this work was obtained by J17 using the {\sc ChaRT} simulated PSF within 15 arcsec and the PSF wing measured by Gaetz (2010) up to 500 arcsec.

\section{Peculiar Eclipse Light Curves of \axj}
Following the data reduction procedures in Section~\ref{sec-data} and based on all the eclipse periods listed in Table~\ref{tab-obs}, we extracted the combined 2-10 keV eclipse light curves of \axj\ from \xmm\ EPIC-pn and \cxo\ ACIS-S, separately. For EPIC-pn, the source extraction region was chosen to be an annular region of 20-100 arcsec in order to avoid the pile-up contamination within 20 arcsec, while for ACIS-S the source extraction region was 2.4-20 arcsec. The background count rate was measured in the same region from previous observations during which \axj\ was in quiescence (see J17), and subtracted from these source light curves. Fig.\ref{fig-lc210}a shows the eclipse light curve observed by EPIC-pn, where a significant excess flux of $\sim$30\% of the non-eclipse flux can be observed during the eclipse phase, and the flux seems to decrease from the ingress to egress time. This excess flux and asymmetric shape of the eclipse light curve is more remarkable when observed by ACIS-S, as shown in Fig.\ref{fig-lc210}b. Right after the ingress time, the flux only drops by $\sim$30\%. Then it keeps decreasing to $\sim$40\% within 0.8 ks. At the egress time, the flux only recovers to $\sim$70\% of the flux before eclipsing, and then it gradually increases to the normal non-eclipse flux at more than 1 ks after the egress time. This clearly suggests that there should be a variable component responding to the source eclipse with a time lag. The difference between the eclipse light curves from \xmm\ and \cxo\ is caused by different PSF sizes and different source extraction regions.

We checked the energy-dependent halo variability by extracting the radial profile of \axj\ in different energy bands for different time intervals during the eclipse phase. For \xmm\ EPIC-pn, we divided the eclipse phase into four time intervals of 300 s each (see Fig.\ref{fig-lc210}a), i.e. for $t=$ 0.15-0.45 ks (Interval 1), 0.45-0.75 ks (Interval 2), 0.75-1.05 ks (Interval 3), 1.05-1.35 ks (Interval 4), and then extracted the radial profile within these intervals for the 2-4, 4-6 and 6-10 keV bands, separately. As a conservative choice, we excluded the first 150 s of the eclipse phase because the flux varies rapidly within this period. Fig.\ref{fig-eclips-rad}a,b,c compare the four eclipse-phase radial profiles of \axj\ and the non-eclipse radial profiles. Although the central 20 arcsec is contaminated by photon pile-up, we plot them for the purpose of comparison. Firstly, the radial profile drops by a factor of few after the ingress time. The flux drop is more dramatic at smaller radii, which is because the PSF dominates at the central region where the time lag of the halo is also smaller. Secondly, the flux drop is more significant at harder X-rays, which is consistent with the energy dependence of the dust scattering opacity (i.e. the dust scattering halo has a smaller distribution at higher energies). Thirdly, there is significant difference between the four eclipse-phase radial profiles within 20 arcsec, where a clear decrease of surface brightness is observed from Interval 1 (blue) to 2 (red), 3 (orange) and 4 (green). Note that photon pile-up affects the higher flux ones more, so this trend of flux drop would be even more significant without the pile-up effect. This is also consistent with the shape of the eclipse light curves in Fig.\ref{fig-lc210}.

For \cxo\ ACIS-S, the number of counts is much smaller than in EPIC-pn, so we extracted the radial profile for the whole eclipse phase in order to achieve the highest signal-to-noise, as shown in Fig.\ref{fig-eclips-rad}d,e,f. There is no similarly large flux drop as observed by \xmm, especially in the 2-4 keV band where the flux drop in ACIS-S is hardly observable at small radii, because the lower energy emission is dominated by the dust scattering component. However, we can still see a larger flux drop at harder X-rays similar to the results of \xmm. This is consistent with the ACIS radial profile decomposition in J17, where the dust scattering halo dominates the flux at large radii due to the small PSF of ACIS, and the halo has a lagged response to the source variability.

\section{Response of the Dust Scattering Halo to the Eclipse Signal}
\label{sec-simu}
\subsection{Simulation Setup}
The significant excess flux during the eclipse phase of \axj\ has been speculated to come from dust scattering for many years (Maeda et al. 1996; Hyodo et al. 2009; Degenaar et al. 2012), but there was no attempt to use dust scattering to model these peculiar eclipse light curves. In the previous section, we reported that the observed eclipse radial profiles and light curves of \axj\ are qualitatively consistent with a lagged response in the dust scattering halo. However, the time lag of a scattered photon can range from zero to hundreds of kilo-seconds depending on the location of the dust layer and the viewing angle, and J17 reported that there should be at least two major dust layers in front of \axj, so it is necessary to understand how the dust scattering halo responds to the eclipse signal for different foreground dust layers.

Compared to previous studies of dust scattering rings around fast X-ray transients (e.g. Beardmore et al. 2016; Heinz et al. 2016), the complexity of studying the eclipse light curve lies in the short eclipse period of \axj\ which is only 4.7\% of the orbital period, as well as the multiple geometrically thick foreground dust layers. Thus it is impossible to directly detect any {\it dark} expanding ring structures in the halo due to the eclipse. Therefore, we first used simulations to understand the response of the halo from different dust layers.

We used the periodic rectangular function in Eq.\ref{equ-4} to approximate the intrinsic eclipse light curve of \axj. We did not model the dips or type-I bursts (Ponti et al. 2017a), because the dips have intrinsically complex structures and are not periodic (so the light curve cannot be folded), and the type-I bursts are not energetic enough for a halo variability study. The source distance was assumed to be 8 kpc\footnote{Since the time lag scales linearly with the source distance (Eq.\ref{equ-1}), putting the source at e.g. 12 kpc only increases all the time lags by a factor of 1.5, which is easily overwhelmed by the time lag variation due to different dust distribution, so the simulation results are not sensitive to the source distance.}. The dust scattering halo was calculated using Eq.\ref{equ-3}, which was then convolved with the instrumental PSF to simulate the observed image at different times. Then the eclipse radial profiles and light curves were extracted from different annular regions. In every simulation, we assumed that there was only one foreground dust layer, and we changed the location of the layer for different simulations. This allows us to compare the halo variability for dust layers located at different distances. We also changed the photon energy in order to investigate the energy dependence of the halo variability.

\begin{figure*}
\begin{tabular}{c}
\includegraphics[bb=140 0 612 710, scale=0.65]{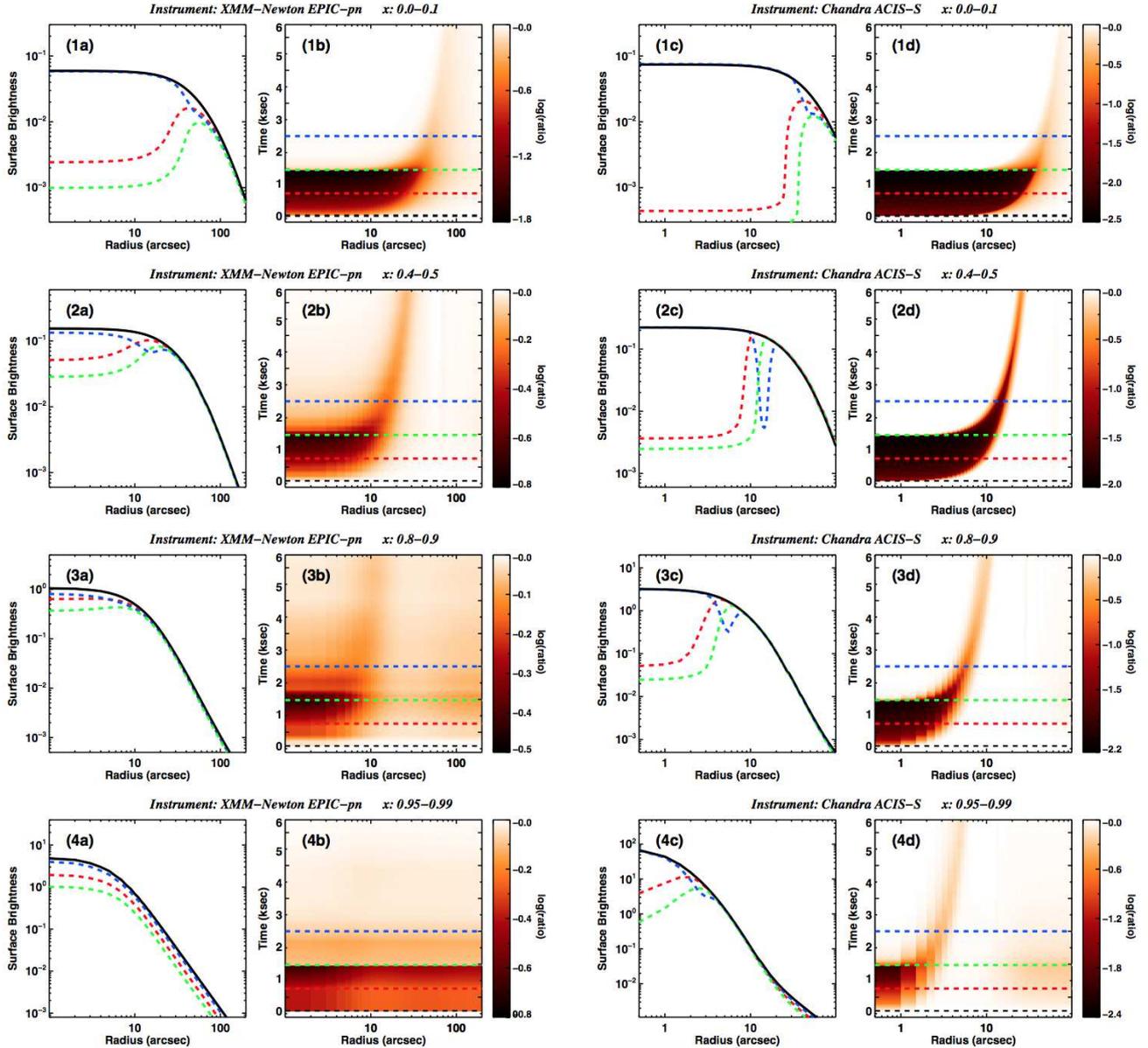} \\
\end{tabular}
\caption{The simulated halo variability induced by the eclipse signal of \axj\ at 5 keV. The four panels in the first row show the halo variability for a dust layer within a fractional distance of 0.0-0.1 from Earth. Panel-1a,1b show the results for \xmm\ EPIC-pn, while Panel-1c,1d are for \cxo\ ACIS-S. In Panel-1a,1c, the black solid line is the non-eclipse halo profile right before the ingress time (i.e. $t=0$ ks). The red, green and blue dash lines are the halo profile at $t=$ 0.7, 1.44 and 2.5 ks. Panel-1b,1d show the halo profile relative to the non-eclipse halo profile at different times after the ingress time, where darker color indicates lower surface brightness. It is clear that the eclipse signal {\it propagates} from inner to outer radii after the ingress time due to longer time lags. The 2nd, 3rd, 4th rows present the results for a dust layer lying within a fractional distance of 0.4-0.5, 0.8-0.9, 0.95-0.99, respectively.}
\label{fig-halotiming}
\end{figure*}

\begin{figure*}
\begin{tabular}{cc}
\includegraphics[bb=10 144 540 600, scale=0.47]{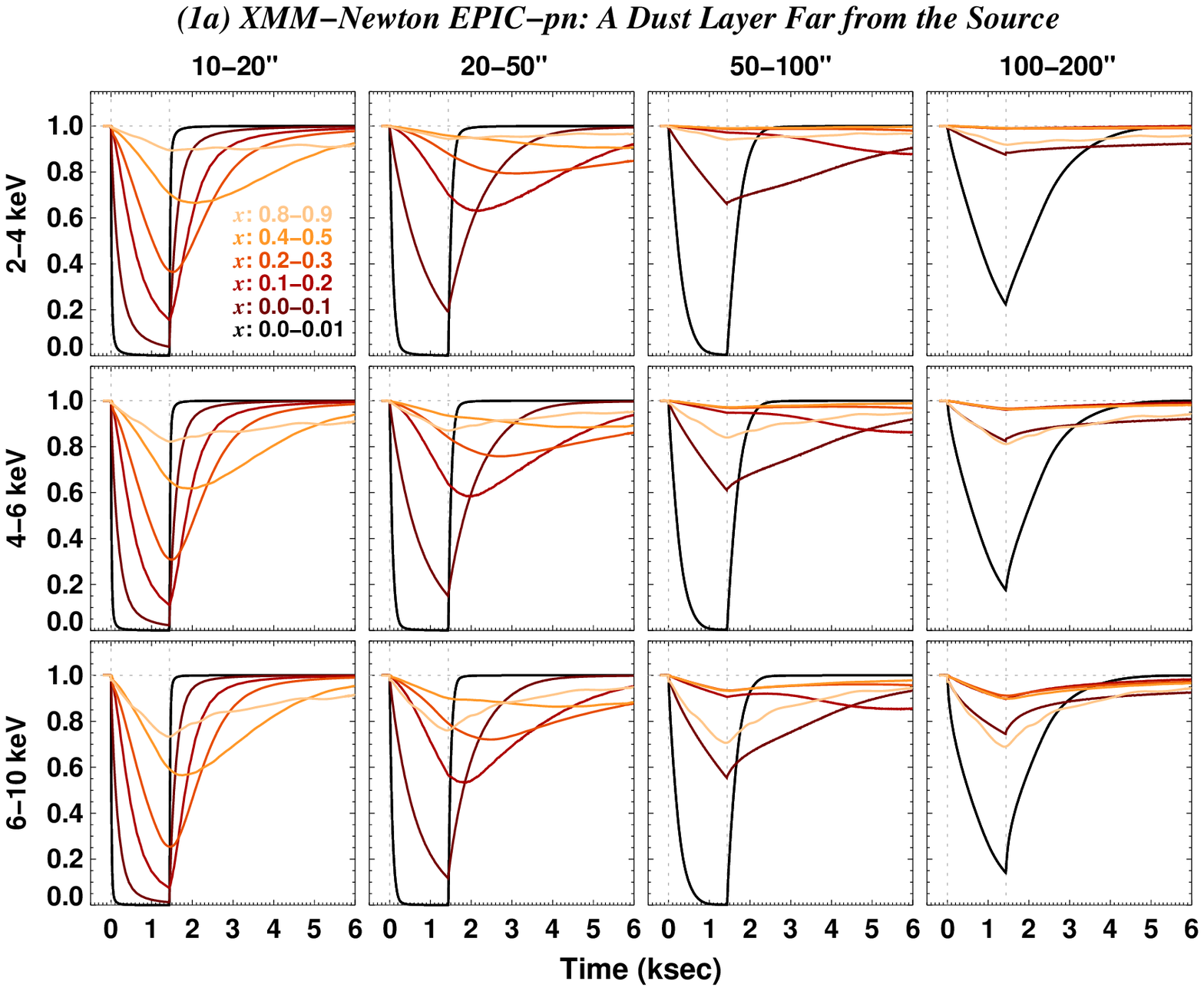} &
\includegraphics[bb=50 144 540 600, scale=0.47]{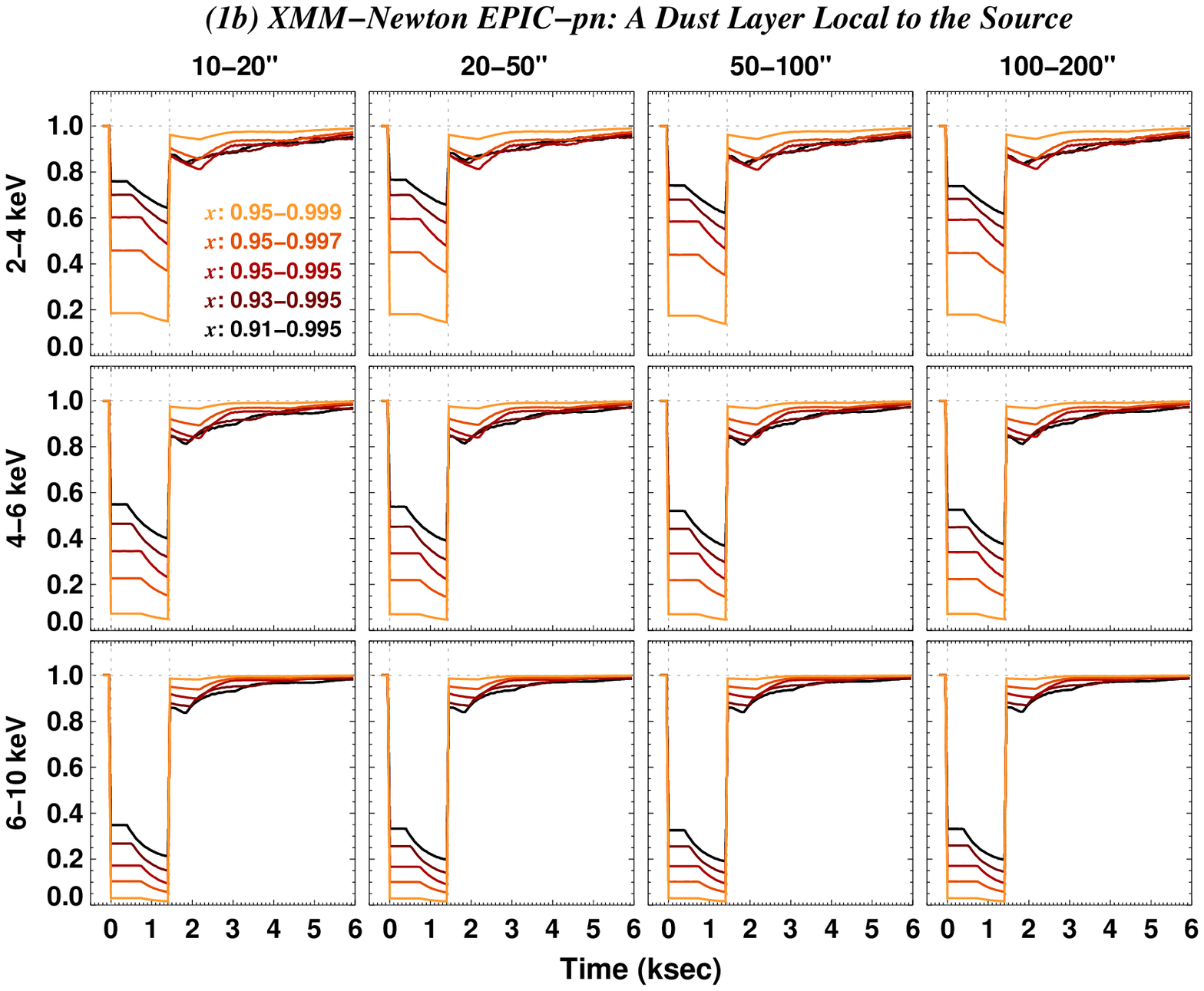} \\
\multicolumn{2}{c}{\includegraphics[bb=35 180 720 350, scale=0.52]{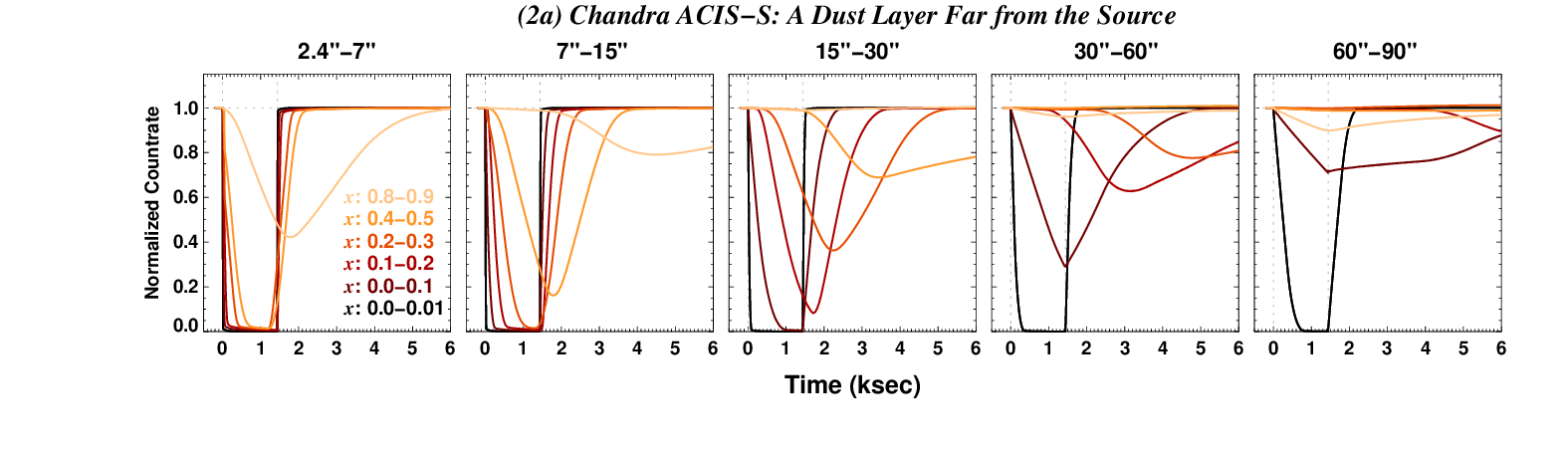}} \\
\multicolumn{2}{c}{\includegraphics[bb=35 180 720 380, scale=0.52]{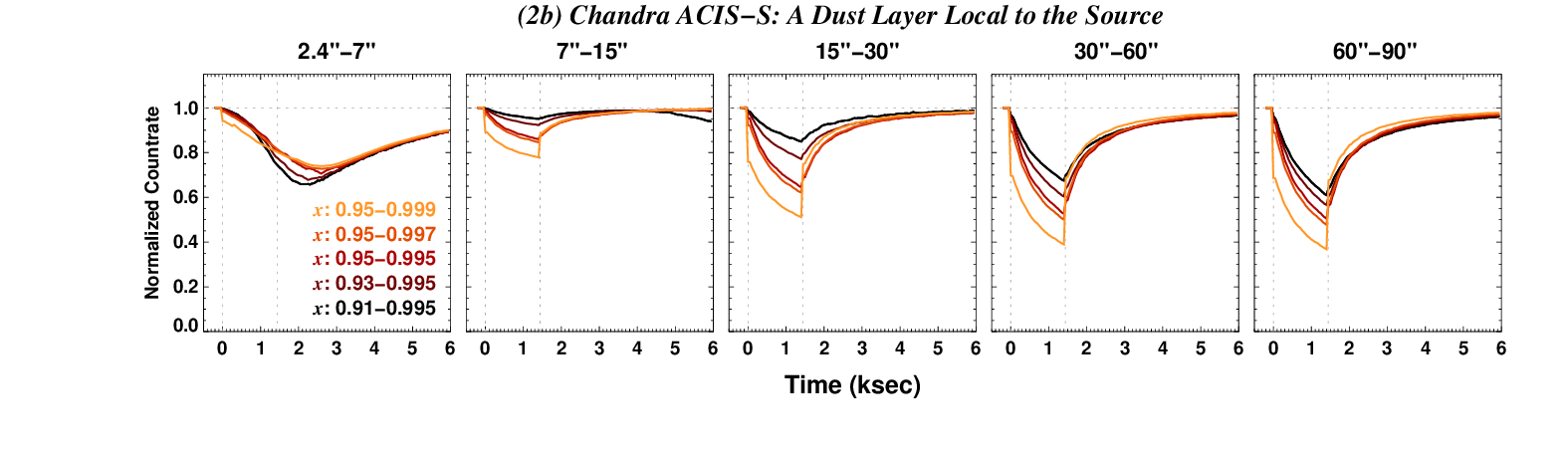}} \\
\end{tabular}
\caption{The simulated eclipse light curves of the dust scattering halo. Panel-1a shows the results for \xmm\ EPIC-pn for a dust layer far from the source. In each sub-panel there are 6 lines showing the results for a dust layer located within a fractional distance of 0.0-0.01, 0.0-0.1, 0.1-0.2, 0.2-0.3, 0.4-0.5, 0.8-0.9, respectively. These sub-panels clearly reveal the dependence of the eclipse light curves on the energy band and source extraction region (i.e. radial range). Panel-1b is similar but for a dust layer local to the source, within a fractional distance of 0.91-0.995, 0.93-0.995, 0.95-0.995, 0.95-0.997, 0.95-0.999, respectively. Panel-2a,2b show the results of simulation for \cxo\ ACIS-S.}
\label{fig-halovar}
\end{figure*}

\subsection{Characteristic Eclipse Radial Profiles and Light Curves of the Dust Scattering Halo}
The simulation was performed for both \xmm\ EPIC-pn and \cxo\ ACIS-S. Fig.\ref{fig-halotiming} shows the simulated halo profiles. Panel-1a shows the results in EPIC-pn for a dust layer distributed smoothly within a fractional distance of 0.0-0.1 from Earth. The black solid line is the halo profile right before the ingress time ($t~=$ 0 ks). The red line shows the radial profile at 0.7 ks after the ingress time, when flux drop is clearly seen at small radii. The flux drop continues until 1.44 ks - the egress time (green dash line) due to the lagged response of the scattering component. After the egress time the source flux recovers to the normal level at small radii. But the radial profile at larger radii does not fully recover even after 2.5 ks (blue dash line), which is also because of the more lagged response. This radial profile variability is better reflected in Panel-1b, where the Y-axis is the time and X-axis is the radius. The color is coded according to the ratio relative to the non-eclipse radial profile, i.e. at time $t \leq 0$, such that a darker color indicates a more dramatic relative flux drop. We can see that the flux within 10 arcsec responds instantaneously to the eclipse signal. The flux within 10-50 arcsec shows a lagged response during the eclipse phase. Then the eclipse signal {\it propagates} to even larger radii after the egress time. Panel-1c,1d shows the result for ACIS-S, where a similar propagation of lagged eclipse signal can be found. The difference between the two instrument is merely due to their different PSFs.

As the dust layer's location becomes closer to the source, the time lag increases at the same viewing angle, and so the lagged eclipse signal can be observed earlier at smaller radii, as shown in the 2nd and 3rd rows in Fig.\ref{fig-halotiming}. However, there is also an interesting change of variability when the dust layer becomes very close to the source. Panel-4a,4b in Fig.\ref{fig-halotiming} show the results in EPIC-pn when the dust layer lies within a fractional distance of 0.95-0.99, i.e. local to the source. In this case, the flux at all radii responds simultaneously at the ingress time, and also shows a gradual decrease during the eclipse phase. This is because the dust layer is too close to the source, then its halo is strongly peaked at the source position, so after the PSF convolution the halo's radial profile is almost identical to the PSF profile, but the halo still has a significant time lag relative to the source PSF. Therefore, The simulated radial profiles in Fig.\ref{fig-halotiming} suggest that the eclipse light curve should show significant differences for different foreground dust distribution, source extraction regions and energy bands. Different instruments will also observe different eclipse light curves due to their different PSFs, which not only determine the radial profile of the central point source, but also change the shape of the observed dust scattering halo.

Fig.\ref{fig-halovar} presents the eclipse light curves for different dust layers. It is clear that a foreground dust layer can easily produce significant excess flux during the eclipse phase and create an asymmetric eclipse light curve. Fig.\ref{fig-halovar} Panel-1a,2a show the results for a dust layer far from the source, where the excess flux increases as the dust layer becomes closer to the source, and the eclipse asymmetry also becomes more remarkable. In comparison, Fig.\ref{fig-halovar} Panel-1b,2b show the results for a dust layer local to the source, where the excess flux decreases as the dust layer becomes closer to the source, which is because of the more peaked dust scattering halo and a faster response of the peak to the eclipse signal.

Fig.\ref{fig-halovar} also shows that the source light curve depends strongly on the source extraction region. At small radii, the flux can be dominated either by the source PSF which responds instantaneously to the eclipse signal, or by the halo from a dust layer very close to the source which shows significant time lag. Thus at small radii the shape of the eclipse light curve can help to constrain the contribution of scattering light from a dust layer local to the source, while at large radii the eclipse light curve can be used to constrain more distant dust layers from the source.

Finally, although these simulations were done for the eclipse signal of \axj, the principle of dust scattering and time lag is the same for all other sources. Therefore, the shape of these eclipse light curves should be typical for any other eclipsing sources with significant amount of LOS dust, and thus can be used to verify the dust scattering origin, such as in Swift J1749.4-2807 (Ferrigno et al. 2011).

\section{Modelling the X-ray Eclipse of \axj}
\subsection{Modelling Strategy}
J17 has modelled the non-eclipse radial profile of \axj\ with two dust layers, one local to the binary system and the other closer to Earth. This nearby thick dust layer likely comprises a few thinner sub-layers in the Galactic disk, but the time-independent radial profile fitting was not able to provide further constraints. Now the simulation in Section~\ref{sec-simu} allows us to understand the response of different LOS dust layers to an eclipse signal, so it is possible to use the eclipsing behaviour of \axj\ to further constrain its foreground dust distribution. The shape of the observed eclipse light curves of \axj\ depends on both photon energy and source extraction region. Besides, the instrumental PSF will also affect the shape of the observed dust scattering halo, so that different instruments will observe different eclipse light curves. {\it Therefore, a good dust scattering model should reproduce all these different light curves.} However, it is not realistic to perform a direct fit to all these eclipse light curves due to the long computing time. Instead, we can perform fitting to all the radial profiles in Fig.\ref{fig-eclips-rad} in parallel. These include 6 non-eclipse radial profiles from EPIC-pn and ACIS-I in 2-4, 4-6 and 6-10 keV bands, plus 12 eclipse radial profiles from EPIC-pn for the four time intervals in Fig.\ref{fig-lc210}a and the three energy bands, plus 3 eclipse radial profiles from ACIS-S for the same three energy bands, and so in total there are 21 radial profiles to be fitted simultaneously.

The difference between the non-eclipse and eclipse radial profiles is due to the eclipsed source PSF and the delayed response of the halo, which can be calculated using Eq.\ref{equ-3}. For modelling the EPIC-pn eclipse radial profiles, we calculated the halo model at $t=300,~600,~900$ and 1200 s; while for the ACIS-S eclipse profiles, we chose the mid-point of the eclipse period at $t=720$ s. We also used the spectral-weighted photon energy, which is 3.3, 5.0 and 7.0 keV for the three energy bands, respectively. J17 also discovered a halo wing component contributing significant flux outside $\sim$200 arcsec in EPIC-pn\footnote{The origin of this halo wing is not clear, but is probably related to an additional dust grain population with smaller typical grain size, see J17.}, so we adopted a free constant to account for this halo wing component at small radii.

Another major uncertainty lies in the variation of the dust grain population along the GC LOS (see Section 1.2 in J17). So far dozens of different dust grain populations have been proposed (e.g. MRN77; Weingartner \& Draine 2001; ZDA04; Xiang et al. 2011), each with different grain compositions and abundances. Fritz et al. (2011) tried various dust grain populations to fit the IR extinction curve in the GC direction. They reported evidence of H$_{2}$O in the ISM dust grains, and so the COMP-AC-S grain population of ZDA04 was recommended for the GC LOS. However, it is also known that the metallicity of the GC is higher than the Galactic disk (e.g. Kubryk, Prantzos \& Athanassoula 2015), and the average dust grain size in the GC may be smaller (Hankins et al. 2017). Thus the property of dust grain population is highly likely to show evolution in the GC direction, and so a single grain population might be too simple for the entire LOS. Besides, there are multiple dust layers along the GC LOS (J17), which can easily lead to thousands of combinations of different grain populations for different layers, so it is not practical to try all of them.

Since our primary objective is to understand the peculiar eclipse light curve of \axj, we decided to try some typical dust grain populations. These include the classic MRN grain population (MRN77), in which the graphite and silicate grains follow a power law size distribution. The BARE-GR-B, COMP-AC-S, COMP-NC-B grain populations are also included, whose dust scattering halos can cover the entire halo shape range produced by all the 15 grain populations in ZDA04\footnote{In the 2-layer scenario in J17, the best-fit $\chi^2$ was from the BARE-GR-B grain population, but the fitting depends on model assumptions, so it does not mean that the other grain populations can be ruled out for the GC LOS.}. We used 2-4 dust layers along the LOS in order to check if the number of adopted dust layers can affect the results. To investigate the possible evolution of dust grain population along the GC LOS, we also tried some special cases where the typical dust grain size is smaller in the layer local to the source (Hankins et al. 2017).

Then based on the best-fit parameters for all the halo profiles, we can simulate various eclipse light curves for different instrumental PSF, energy bands and source extraction regions, and compare them to the observed ones as a {\it posteriori} check. For EPIC-pn, we extracted folded eclipse light curves from the annuli of 10-20, 20-50, 50-100 and 100-200 arcsec in the 2-4, 4-6 and 6-10 keV bands. Since the 10-20 arcsec annulus should be slightly contaminated by the photon pile-up effect, we use the result in this region for comparison, but did not include it in the statistics. We did not consider the data outside 200 arcsec because it is dominated by the halo wing component, whose origin is yet unknown (see J17). For ACIS-S, the folded eclipse light curves were extracted from the annuli of 2.4-7, 7-15, 15-30, 30-60, 60-90 arcsec for the entire 2-10 keV band in order to achieve better signal-to-noise.

We emphasize that a reasonable fit to all the halo profiles cannot guarantee a perfect match to all the eclipse light curves. This is because our halo model adopted some approximations and assumptions for simplicity (see Section~\ref{sec-uncertainty}). Another issue is the variability of the halo wing component, as its location and origin are not clear, but it can certainly affect the simulated eclipse light curves at large radii. So we tried two extreme cases, one assuming that the wing does not respond to the eclipse signal, and the other assuming that it responds instantaneously to the eclipse signal. For a source at 8 kpc away and a foreground dust layer at $x=0.1$, the time lag at a viewing angle of 200-300 arcsec is 43-97 ks (i.e. much longer than the eclipse period of 1.44 ks), while for $x=0.001$ the time lag at 200-300 arcsec is 0.39-0.87 ks. Thus the immediate response of the halo wing to the eclipse signal can happen only if the dust layer is merely located a few parsecs away from Earth. However, we consider this possibility unlikely. Indeed, there is evidence that the Solar system is located inside an ISM cavity of 80-200 pc wide (e.g. Tanaka \& Bleeker 1977; see Section~\ref{sec-dsn}). Thus $x=0.001$ is only used for the model tests.

\begin{figure*}
\begin{tabular}{ll}
\includegraphics[bb=5 -400 812 500, scale=0.18]{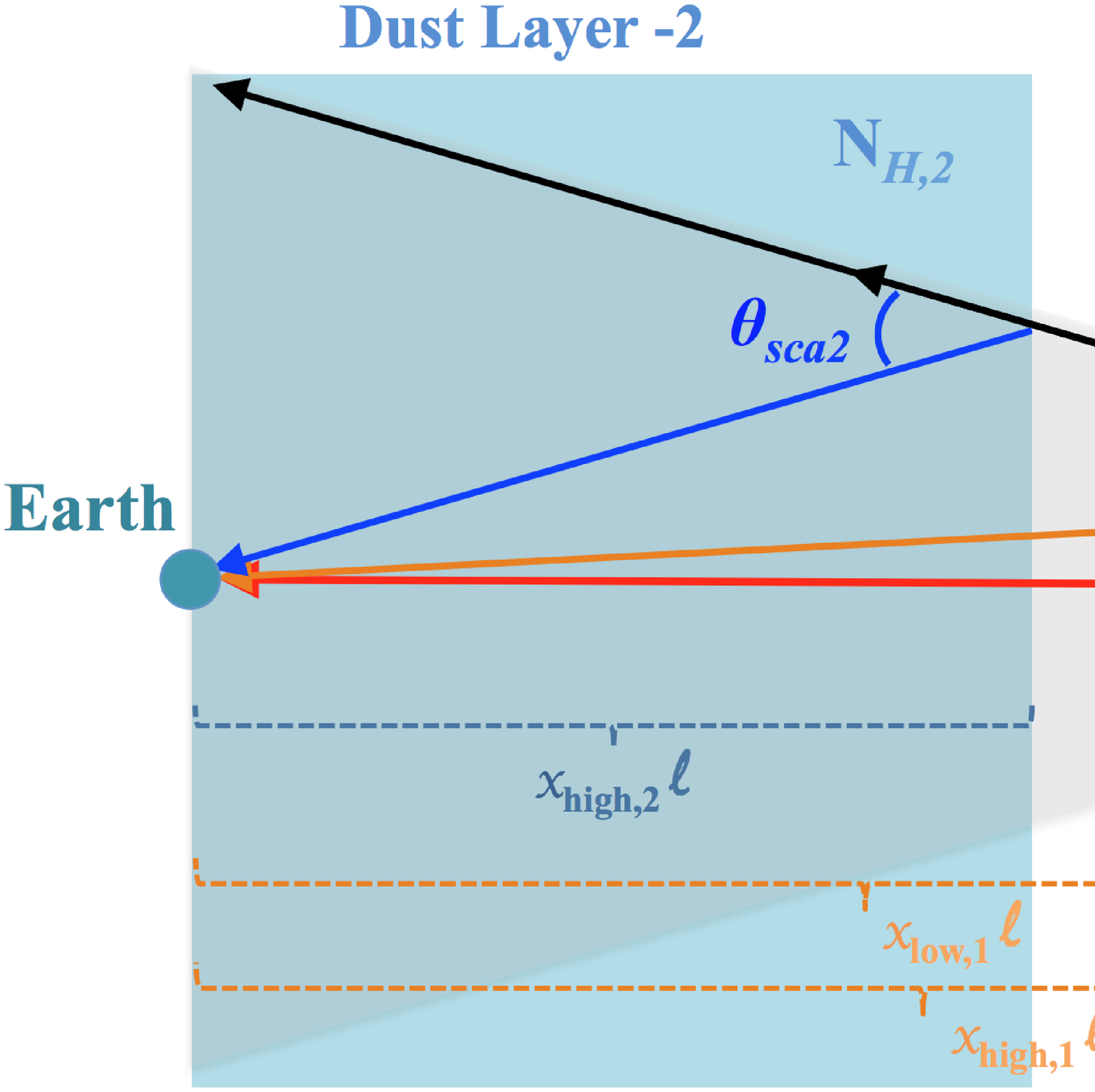} &
\includegraphics[bb=-5 230 666 648, scale=0.47]{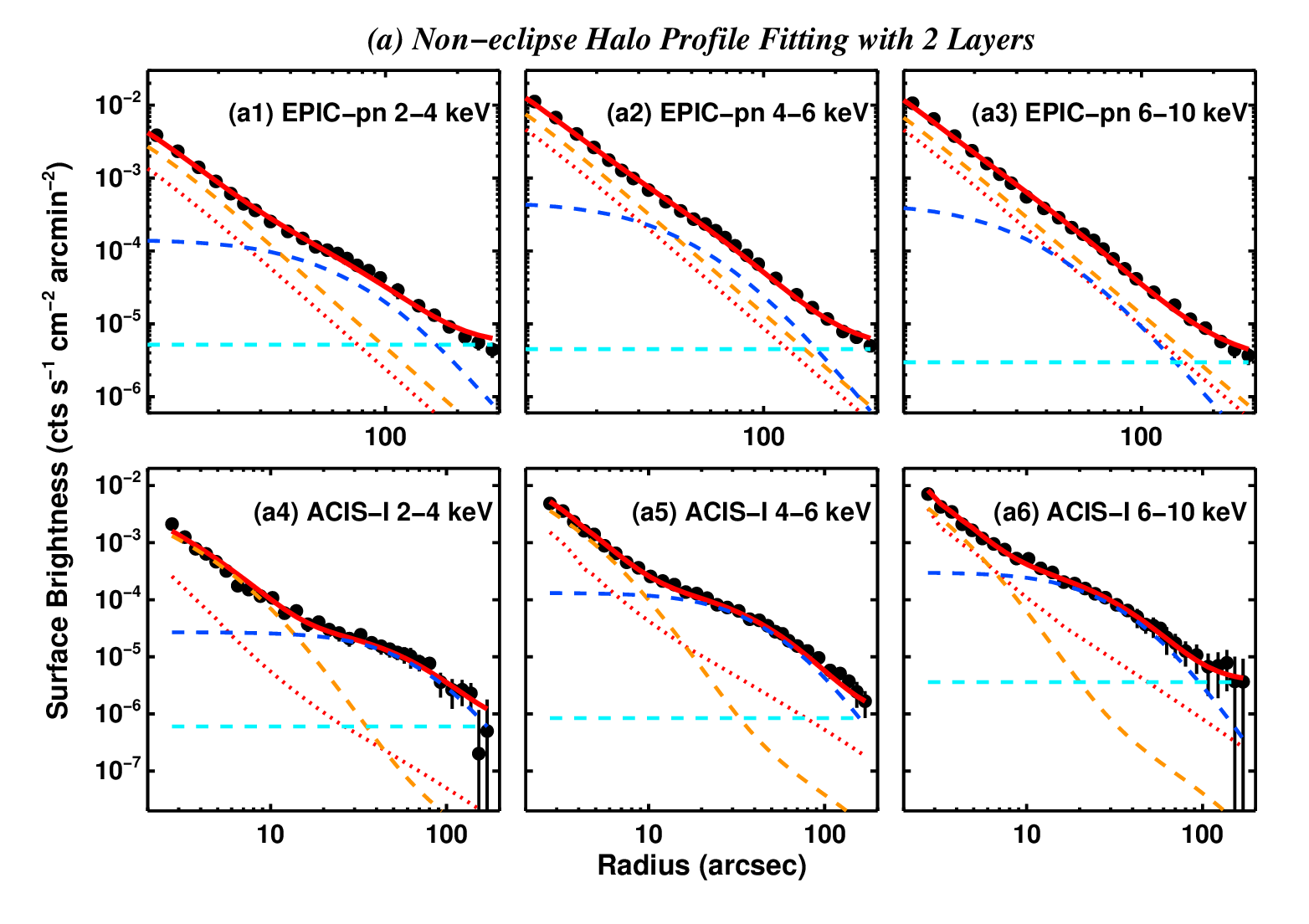} \\
\includegraphics[bb=5 -350 812 500, scale=0.18]{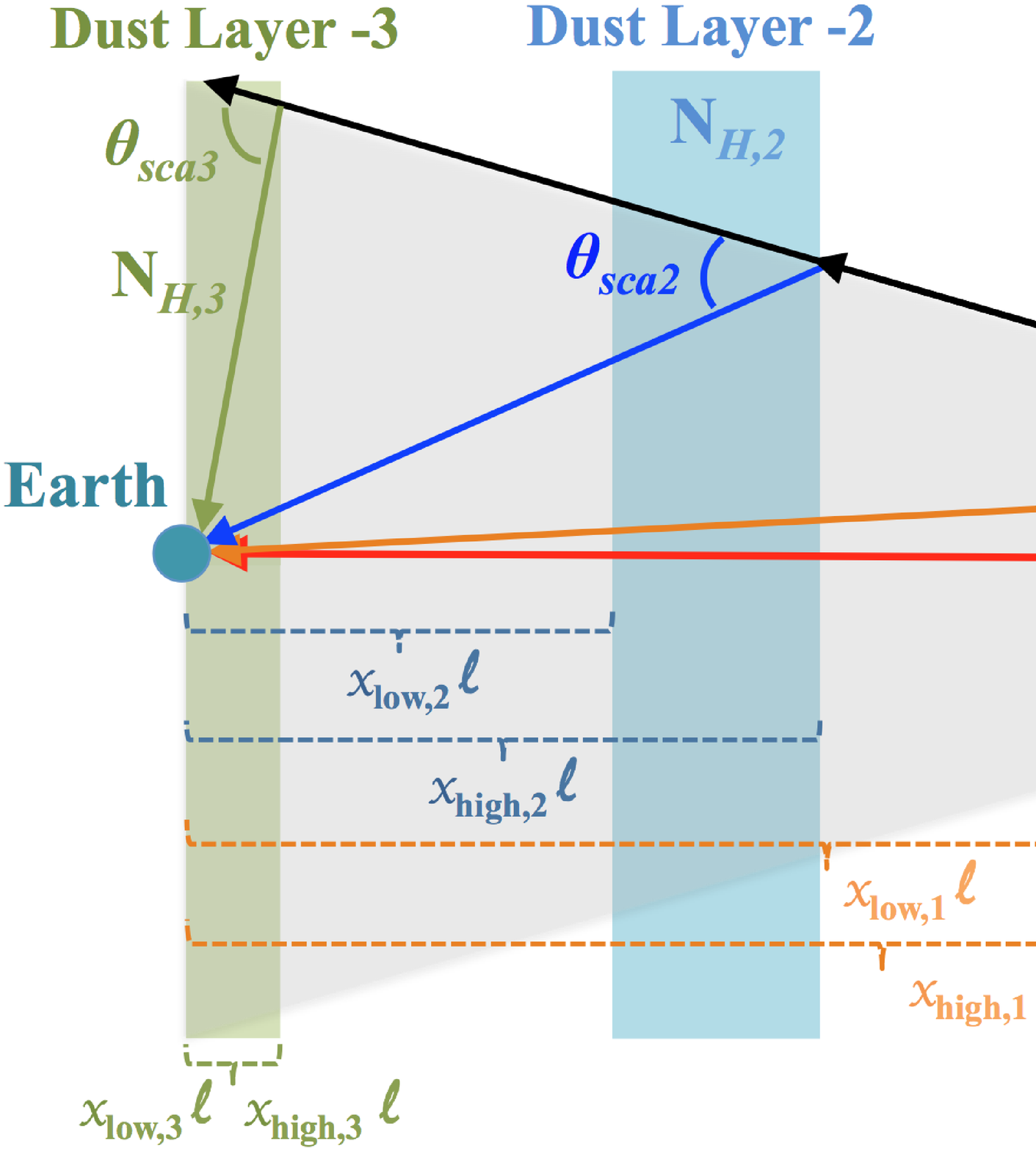} &
\includegraphics[bb=-5 230 666 648, scale=0.47]{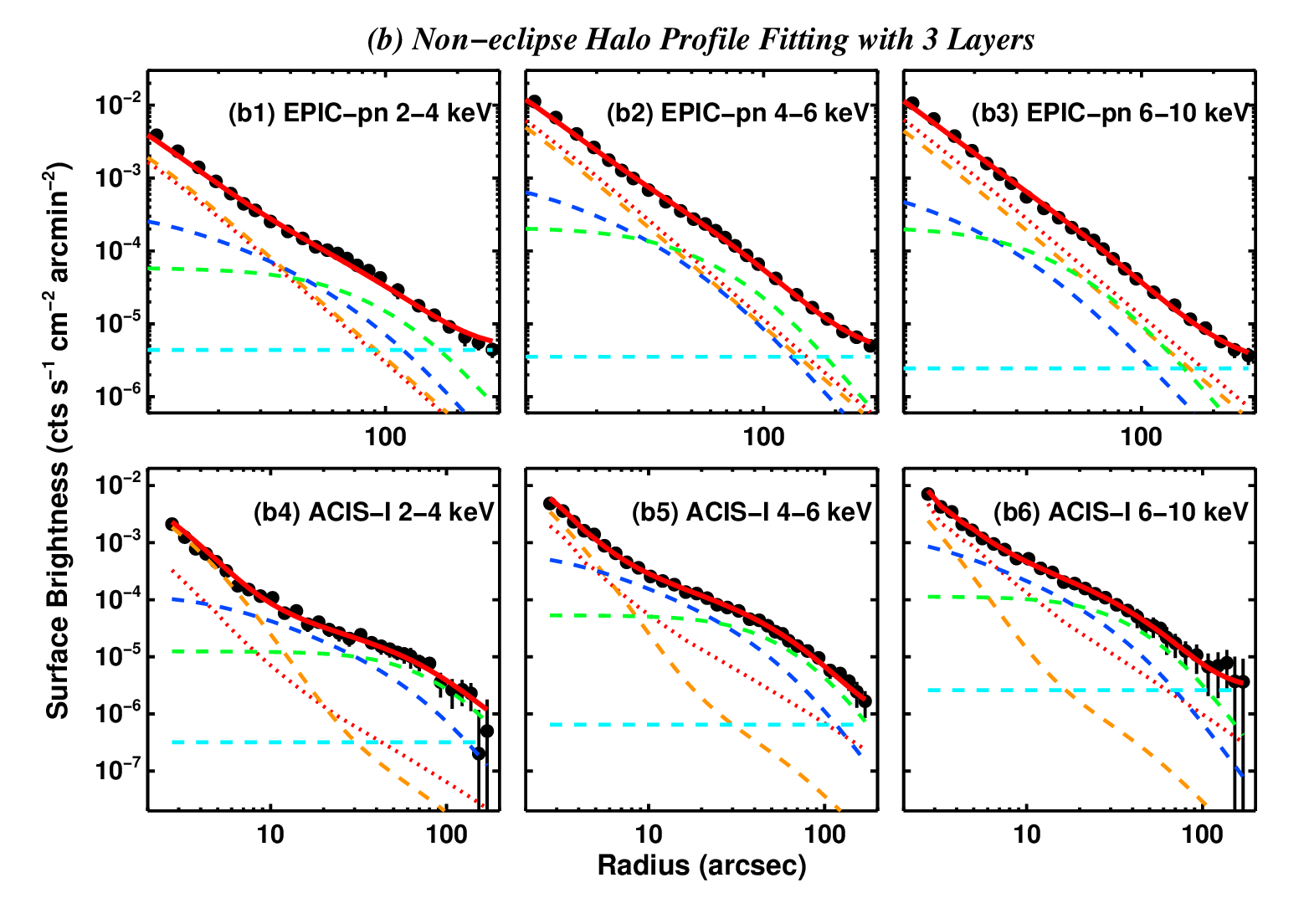} \\
\includegraphics[bb=5 -280 812 500, scale=0.211]{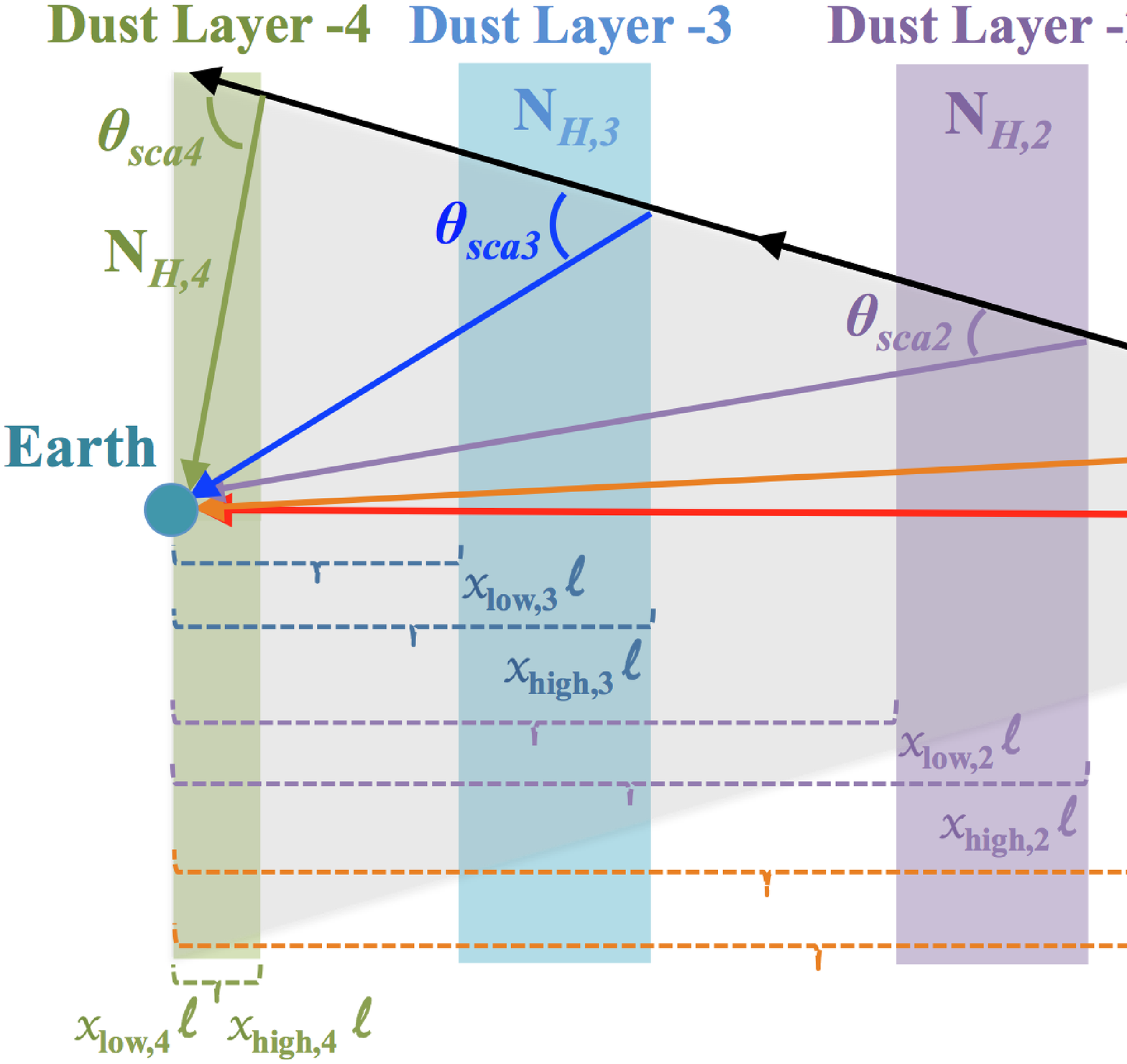} &
\includegraphics[bb=-5 230 666 648, scale=0.47]{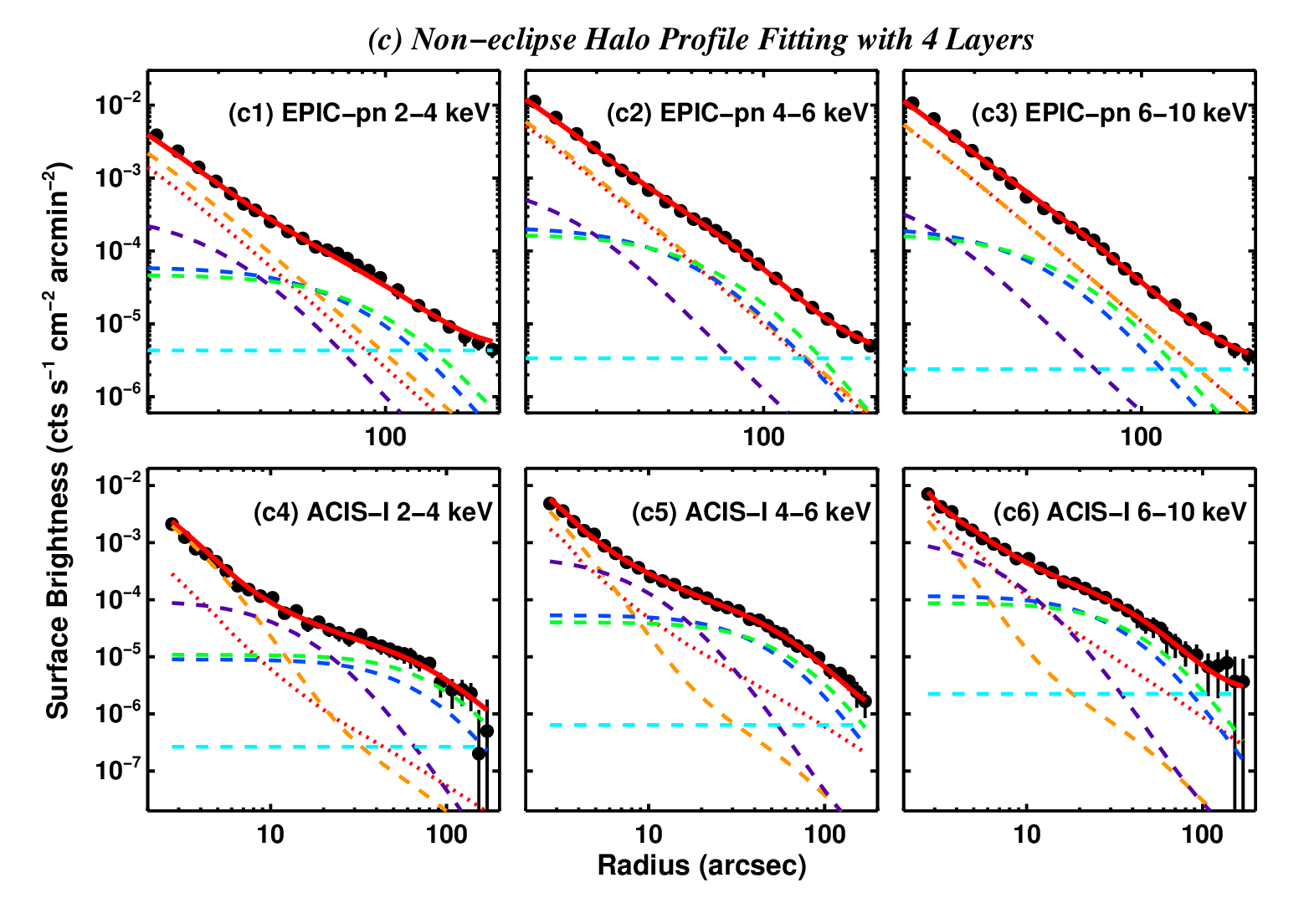} \\
\end{tabular}
\caption{Dust models with 2, 3 and 4 distinct layers along the LOS. In each row, the left-side cartoon shows each layer's position and parameters. The best-fit results using the COMP-AC-S dust grain model (see Section~\ref{sec-fit-m10}) are shown on the right side, with dash lines indicating the contribution from every dust scattering component (cyan dash line is the wing component), and the dotted line indicating the instrumental PSF. The results of simultaneous fitting to the eclipse halo profiles using the same model can be found in Fig.~\ref{app-fig-fit-eclips-2p}-\ref{app-fig-fit-eclips-4p}.}
\label{fig-mlayers}
\end{figure*}

\begin{figure*}
\begin{tabular}{l}
\includegraphics[bb=-50 216 738 750, scale=0.55]{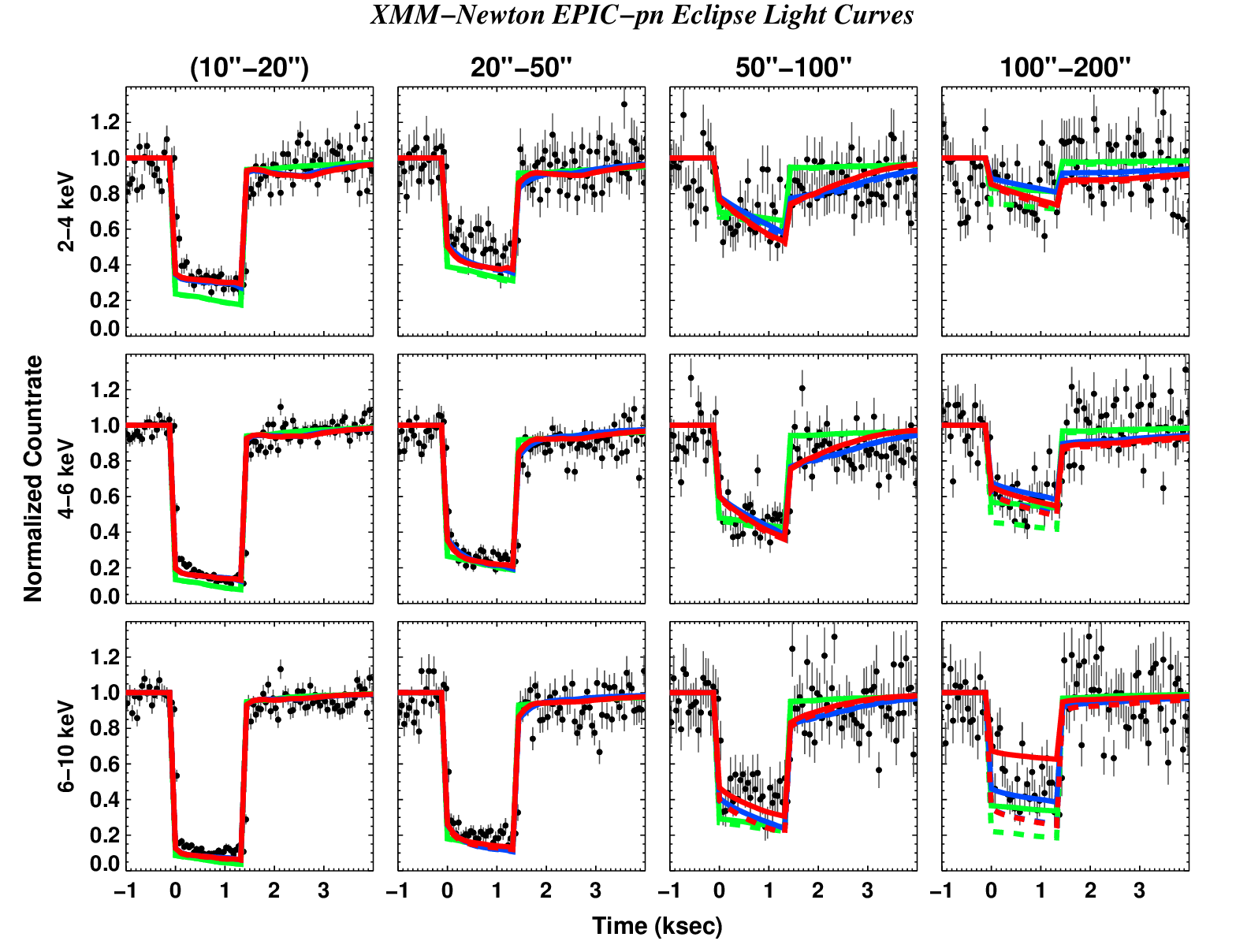} \\
\includegraphics[bb=54 216 954 470, scale=0.55]{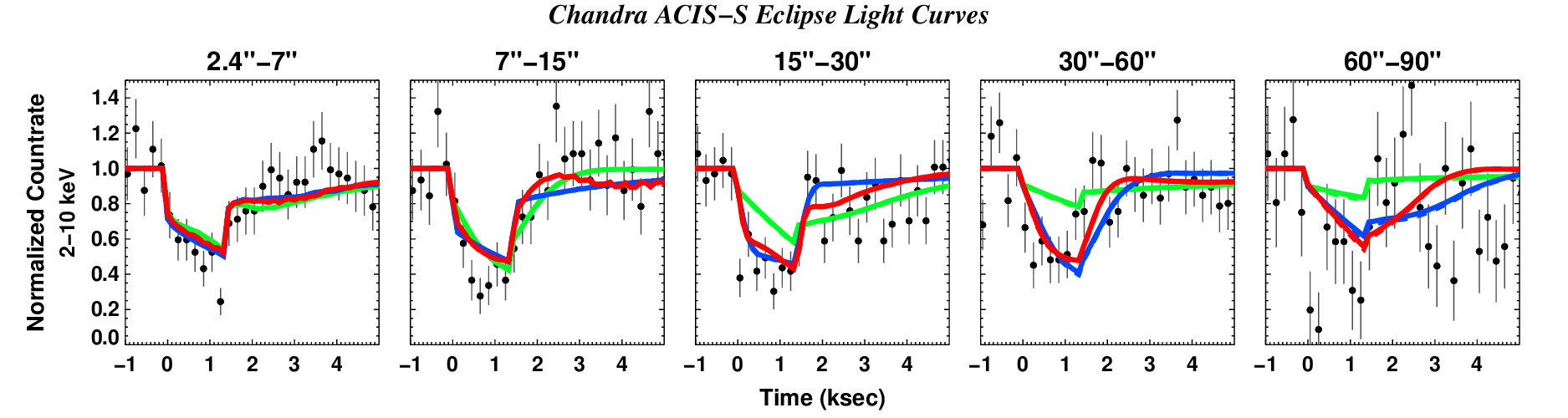} \\
\end{tabular}
\caption{Posterior comparison between the real eclipse light curves observed by \xmm\ EPIC-pn and \cxo\ ACIS-S, and the simulated eclipse light curves from the best-fit models with the COMP-AC-S dust grain model for different dust-layer scenarios (Fig.\ref{fig-mlayers}, Table~\ref{tab-bestfit-m10}). Green, blue and red lines indicate the results for the 2, 3 and 4 dust-layer scenarios, respectively. The solid line assumes that the halo wing does not respond to the eclipse signal, while the dash line assumes that the halo wing responds instantaneously to the eclipse signal. In EPIC-pn the 10-20 arcsec region was slightly contaminated by the photon pile-up effect, so we only plot them for the purpose of comparison. The results for the other dust grain models tried in this paper can be found in Fig.\ref{app-fig-eclips-GM2}-\ref{app-fig-eclips-GM4}.}
\label{fig-eclips-m10}
\end{figure*}

\begin{figure}
\includegraphics[bb=5 15 612 575, scale=0.435]{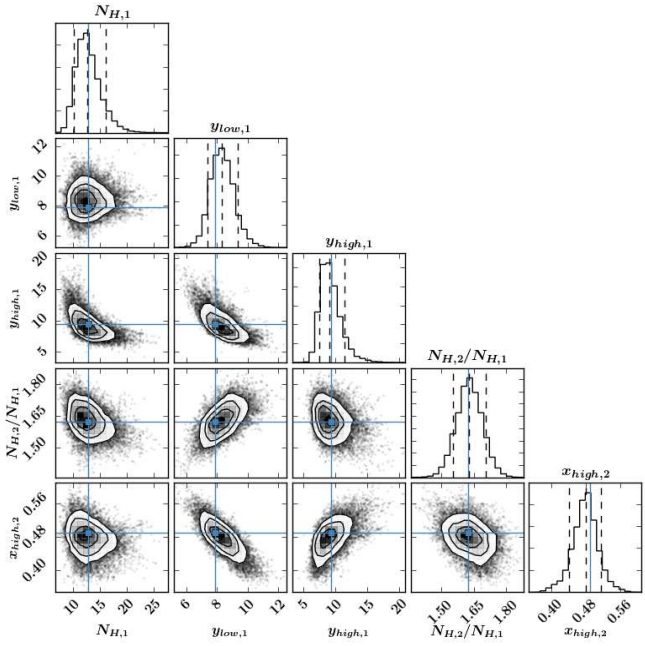} \\
\caption{The bayesian parameter estimation using MCMC sampling for the 2-layer scenario (Fig.\ref{fig-mlayers} 1st row) with the COMP-AC-S dust grain model. Each histogram shows the marginalized distribution, the three dash lines from left to right indicate the 10th, 50th and 90th percentiles of the samples, respectively. For clarity, we performed the linear conversion: $y_{\rm low,1}=(1-x_{\rm low,1})\times10^{2}$, $y_{\rm high,1}=(1-x_{\rm high,1})\times10^{4}$. The blue solid lines indicate the best-fit values for the minimum $\chi^2$ method (see Table~\ref{tab-bestfit-m10}). The results for 3 and 4 dust layers are shown in Fig.\ref{app-fig-mcmc-3p}-\ref{app-fig-mcmc-4p}.}
\label{fig-mcmc-2p}
\end{figure}

\setlength\dashlinedash{0.6pt}
\setlength\dashlinegap{1.5pt}
\setlength\arrayrulewidth{0.8pt}
\begin{table}
\centering
\caption{Results of modelling the halo radial profiles with the GM1 dust grain configuration (i.e. using the COMP-AC-S dust grain model for all the layers). We show both the best-fit parameter values using the minimum $\chi^2$ method and the bayesian parameter estimation using the MCMC sampling. The bayesian values are for the 50th percentiles of the samples, with upper and lower uncertainties indicating the 10th and 90th percentiles (see Fig.\ref{fig-mcmc-2p}, Fig.\ref{app-fig-mcmc-3p}, \ref{app-fig-mcmc-4p}). $f_{\rm nH, m}$ indicates the percentage of dust contained in layer $m$. $N_{\rm H,tot}$ is the total LOS $N_{\rm H}$ towards the source. The meaning of the other parameters are shown in Fig.\ref{fig-mlayers}.}
\begin{tabular}{llcccc}
\hline
& Parameter & Value-$\chi^2_{\rm \nu}$ & Value-Bayesian & Unit \\
\hline
\multicolumn{2}{l}{2-layer Scenario} & & \\ 
\hdashline
Layer-1 & x$_{\rm low,1}$ & 0.921 & 0.917$^{+0.009}_{-0.010}$ & \\
& x$_{\rm high,1}$ & 0.999 & 0.991$^{+0.002}_{-0.002}$ & \\
& $f_{\rm nH,1}$ & 38.1 & 38.0$^{+1.2}_{-1.1}$ & \% \\
\hdashline
Layer-2 & x$_{\rm low,2}$ & 0-fixed & 0-fixed & \\
& x$_{\rm high,2}$ & 0.490 & 0.479$^{+0.036}_{-0.039}$ \\
& $f_{\rm nH,2}$ & 61.9 & 62.0$^{+1.1}_{-1.2}$ & \% \\
\hdashline
& $N_{\rm H,tot}$ & 33.6 & 33.1$^{+10.3}_{-7.0}$ & $10^{22}$ cm$^{-2}$\\
\hline
\multicolumn{2}{l}{3-layer Scenario} & & \\
\hdashline
Layer-1& x$_{\rm low,1}$ & 0.966 & 0.967$^{+0.004}_{-0.003}$ &\\
& x$_{\rm high,1}$ & 0.996 & 0.994$^{+0.002}_{-0.004}$ & \\
& $f_{\rm nH,1}$  & 37.5 & 37.3$^{+8.3}_{-7.5}$ & \% \\
\hdashline
Layer-2& x$_{\rm low,2}$ & 0.116 & 0.221$^{+0.109}_{-0.093}$ &\\
& x$_{\rm high,2}$ & 0.947 & 0.933$^{+0.014}_{-0.023}$ &\\
& $f_{\rm nH,2}$ & 26.6 & 24.0$^{+3.9}_{-3.5}$ & \% \\
\hdashline
Layer-3 & x$_{\rm low,3}$ & 0-fixed & 0-fixed & \\
& x$_{\rm high,3}$ & 0.058 & 0.068$^{+0.012}_{-0.012}$& \\
& $f_{\rm nH,3}$ & 35.9 & 38.7$^{+4.6}_{-3.7}$ & \% \\
\hdashline
& $N_{\rm H,tot}$ & 33.3 & 29.6$^{+9.7}_{-7.6}$& $10^{22}$ cm$^{-2}$\\
\hline
\multicolumn{2}{l}{4-layer Scenario} & & \\
\hdashline
Layer-1 & x$_{\rm low,1}$ & 0.970 & 0.968$^{+0.004}_{-0.003}$ & \\
& x$_{\rm high,1}$ & 0.995 & 0.996$^{+0.001}_{-0.001}$ & \\
& $f_{\rm nH,1}$ & 40.4 & 38.6$^{+8.0}_{-6.8}$ & \% \\
\hdashline
Layer-2 & x$_{\rm low,2}$ & 0.840 & 0.910$^{+0.031}_{-0.063}$& \\
& x$_{\rm high,2}$ & 0.840 & 0.910$^{+0.031}_{-0.063}$ &  \\
& $f_{\rm nH,2}$ & 9.0 & 10.9$^{+2.9}_{-2.1}$ & \% \\
\hdashline
Layer-3 & x$_{\rm low,3}$ & 0.104 & 0.179$^{+0.105}_{-0.068}$& \\
& x$_{\rm high,3}$ & 0.384 & 0.313$^{+0.191}_{-0.082}$& \\
& $f_{\rm nH,3}$  & 23.7 & 19.0$^{+4.6}_{-5.9}$ & \%\\
\hdashline
Layer-4 & x$_{\rm low,4}$ & 0-fixed & 0-fixed & \\
& x$_{\rm high,4}$ & 0.035 & 0.045$^{+0.023}_{-0.012}$& \\
& $f_{\rm nH,4}$  & 26.9 & 31.6$^{+4.8}_{-4.3}$ & \%\\
\hdashline
& $N_{\rm H,tot}$  & 39.0 & 39.5$^{+14.1}_{-10.9}$ & $10^{22}$ cm$^{-2}$\\
\hline
\end{tabular}
\label{tab-bestfit-m10}
\end{table}

\begin{table}
 \centering
  \centering
   \caption{Dust grain configurations indicated by different model names. The position of every layer is shown in Fig.~\ref{fig-mlayers}. MRN is the dust grain population reported in Mathis, Rumpl \& Nordsieck (1997). COMP-AC-S, COMP-NC-S and BARE-GR-B are the dust grain populations reported in Zubko, Dwek \& Arendt (2004).}
    \label{tab-modelnames}
     \begin{tabular}{@{}ccccc@{}}
     \hline
     Model & Layer-1 & Layer-2 & Layer-3 & Layer-4 \\
     \hline
     2L-GM1 & COMP-AC-S & COMP-AC-S & -- & -- \\
     3L-GM1 & COMP-AC-S & COMP-AC-S & COMP-AC-S & -- \\
     4L-GM1 & COMP-AC-S & COMP-AC-S & COMP-AC-S & COMP-AC-S \\
     \hline
     2L-GM2 & MRN & MRN & -- & -- \\
     3L-GM2 & MRN & MRN & MRN & -- \\
     4L-GM2 & MRN & MRN & MRN & MRN \\
     \hline
     2L-GM3 & COMP-AC-S & COMP-NC-B & -- & -- \\
     3L-GM3 & COMP-AC-S & COMP-NC-B & COMP-NC-B & -- \\
     4L-GM3 & COMP-AC-S & COMP-NC-B & COMP-NC-B & COMP-NC-B \\
     \hline
     2L-GM4 & BARE-GR-B & COMP-NC-B & -- & -- \\
     3L-GM4 & BARE-GR-B & COMP-NC-B & COMP-NC-B & -- \\
     4L-GM4 & BARE-GR-B & COMP-NC-B & COMP-NC-B & COMP-NC-B \\
     \hline
     \end{tabular}
\end{table}

\setlength\dashlinedash{0.6pt}
\setlength\dashlinegap{1.5pt}
\setlength\arrayrulewidth{0.8pt}
\begin{table}
\centering
\caption{$\chi^2$ statistics for different models in Table~\ref{tab-modelnames}. RP$_{\rm non-eclipse}$ indicates results for the fitting of non-eclipse halo profiles, while RP$_{\rm eclipse}$ is for the fitting of eclipse halo profiles. LC$_{\rm pn}$ and LC$_{\rm ACIS}$ are the {\it posterior} comparison between the simulated eclipse light curves from the best-fit halo models and the observed ones for \xmm\ pn and \cxo\ ACIS. 2L-GM1 signifies 2-layer model with the GM1 dust grain configuration and a constant halo wing (see Section~\ref{sec-fit-m10}).  2L-GM1-wing indicates a similar model as 2L-GM1 but with a halo wing reacting to the eclipse signal instantaneously. The same convention applies to the other model names.}
\begin{tabular}{@{}lcccccc@{}}
\hline
Model & \multicolumn{2}{c}{RP$_{\rm non-eclipse}$} & \multicolumn{2}{c}{RP$_{\rm eclipse}$} & LC$_{\rm pn}$ & LC$_{\rm ACIS}$\\
\hline
& $\chi^2$ & d.o.f & $\chi^2$ & d.o.f & $\chi^2$ & $\chi^2$ \\
\hline 
2L-GM1 & 326.3 & 131 & 763.1 & 352 & 5636.7 & 266.8 \\
3L-GM1 & 198.7 & 128 & 533.2 & 352 & 4599.6 & 218.5 \\
4L-GM1 & 190.5 & 125 & 528.7 & 352 & 4786.1 & 206.6 \\
\hline 
2L-GM1-wing & -- & -- & -- & -- & 5969.7 & 263.3 \\
3L-GM1-wing & -- & -- & -- & -- & 4726.2 & 217.8 \\
4L-GM1-wing & -- & -- & -- & -- & 4860.5 & 206.0 \\
\hline 
2L-GM2 & 318.4 & 131 & 723.3 & 352 & 5693.0 & 281.0 \\
3L-GM2 & 198.2 & 128 & 654.7 & 352 & 5421.3 & 269.9 \\
4L-GM2 & 199.5 & 125 & 649.8 & 352 & 5423.4 & 265.0 \\
\hline 
2L-GM3 & 414.3 & 131 & 773.3 & 352 & 6746.3 & 286.4 \\
3L-GM3 & 257.9 & 128 & 626.8 & 352 & 4455.6 & 226.2 \\
4L-GM3 & 247.7 & 125 & 601.0 & 352 & 4661.0 & 202.7 \\
\hline 
2L-GM4 & 416.6 & 131 & 841.1 & 352 & 5023.7 & 269.7 \\
3L-GM4 & 243.9 & 128 & 619.7 & 352 & 4759.6 & 226.2 \\
4L-GM4 & 236.0 & 125 & 618.9 & 352 & 4849.6 & 260.1 \\
\hline
\end{tabular}
\label{tab-chi2}
\end{table}

\begin{figure}
\centering
\includegraphics[bb=65 160 594 684, scale=0.46]{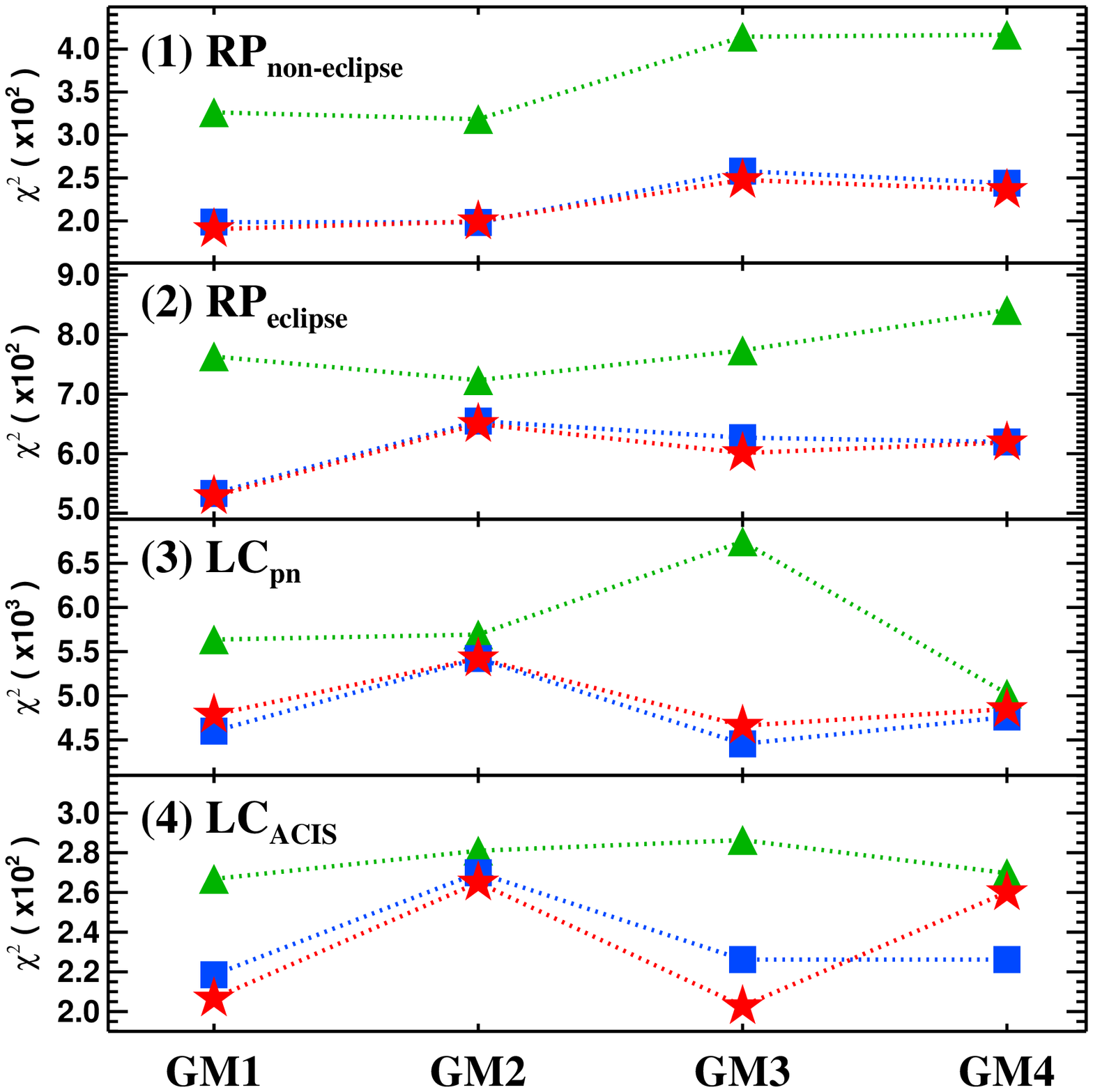}
\caption{Comparison of different $\chi^2$ values in Table~\ref{tab-chi2}. The green triangle, blue square and red star symbols indicate the results for the 2, 3 and 4 dust-layer scenarios, respectively.}
\label{fig-chi2}
\end{figure}

\subsection{Results of the Modelling}
\label{sec-results}
\subsubsection{Halo Models with the COMP-AC-S Dust Grain Population}
\label{sec-fit-m10}
As shown by the cartoons in Fig.~\ref{fig-mlayers}, we tried three scenarios with 2, 3 and 4 dust layers, separately. In the 2-layer scenario, one layer is close to the source while the other is close to Earth. There are 1-2 additional dust layers on the half way in the 3- and 4-layer scenarios. Since the COMP-AC-S grain population was the only one recommended for the GC LOS (Fritz et al. 2011), we started our analysis by adopting it for every dust layer (hereafter: the GM1 grain configuration, see Table~\ref{tab-modelnames}). The best-fit results can be found in Fig.~\ref{fig-mlayers},~\ref{app-fig-fit-eclips-2p}-\ref{app-fig-fit-eclips-4p} and Table~\ref{tab-bestfit-m10} for all three scenarios.

In the 2-layer scenario, the halo decomposition and the best-fit parameters are very similar to the set of values reported in J17, with Layer-1 being local to the source and containing 38.1\% of the LOS dust, while the remaining dust is in Layer-2 distributing from Earth to a fractional distance of 0.490. We notice that the best-fit values are not exactly the same as J17, and the $\chi^2$ of fitting the non-eclipse radial profiles is also slightly worse than in J17. This is because J17 was focusing on the non-eclipse halo profile fitting; while in this work we fitted both non-eclipse and eclipse halo profiles, therefore adding information and further constraining the fit. Since the parameter space can be very complicated and may contain various degeneracies, we used the Bayesian parameter estimation together with the Markov Chain Monte Carlo (MCMC) sampling to explore the parameter space\footnote{We used the Python {\sc emcee} package to perform the Bayesian analysis (Hogg, Bovy \& Lang 2010).}. As shown in Fig.\ref{fig-mcmc-2p} and Table~\ref{tab-bestfit-m10}, our Bayesian analysis found very similar results as the minimum $\chi^2$ method, and the marginalized distributions are all narrow and single-peaked.

Then we simulated the eclipse light curves with the set of best-fit values and compared them to the observation. In Fig.\ref{fig-eclips-m10}, we show that the 2-layer model (green solid lines) can reproduce the energy, radial and instrumental dependences of the halo's eclipse light curves\footnote{This is for the case when the halo wing is not varying. We also tried the case when the halo wing responses to the eclipse signal instantaneously, but only found worse match to the observation, as shown by the green dash lines in the same figure.}. This is a strong confirmation that the 2-layer dust scattering model in both this work and J17 captures the essence of the eclipse phenomena in \axj. However, there are still some significant discrepancies between the model and the data. For instance, the top left panel, displaying the EPIC-pn 2-4 keV light curves at 10-50 arcsec, shows that the simulated eclipse is deeper than the observation. This implies that Layer-1 should have more contribution at inner radii, because its delay time is longer (see Fig.\ref{fig-halovar}). On the contrary, the panels displaying the ACIS 15-90 arcsec light curves show that the simulated eclipse is shallower than the observation, implying that part of the dust scattering happens closer to Earth than the model. Therefore, it is necessary to introduce more dust layers.

In Fig.\ref{fig-mlayers} and Table~\ref{tab-bestfit-m10} we show the best-fit halo decomposition for the 3-layer scenario. Now Layer-2 has a fractional distance of 0.116-0.947 with 26.6\% LOS dust. Layer-3 is within a fractional distance of 0.058 to Earth and contains 35.9\% LOS dust, so its halo shape is most extended and responses most rapidly to the eclipse signal. Layer-1 is local to \axj\ as in the 2-layer best-fit model. The Bayesian analysis found similar results as the minimum $\chi^2$ method (Table~\ref{tab-bestfit-m10}, Fig.\ref{app-fig-mcmc-3p}). Note that for the boundaries of Layer-2, some MCMC steps reached the prior limits corresponding to the fractional distance range of 0.10-0.95. This range was adopted as a prior in the model in order to prevent overlapping of different layers, so that all the dust below the fractional distance of 0.1 (or above 0.95) become part of Layer-3 (Layer-1). Since this prior distance range is already very wide, enlarging it a bit more would only slightly increase the Bayesian errors for the parameters of Layer-2, which would still indicate consistency with the results from the minimum $\chi^2$ method. Compared to the 2-layer best-fit model, we found the $\chi^2$ improves by 127.6 for 3 more degrees of freedom (d.o.f) for non-eclipse radial profiles and by 230 for eclipse radial profiles, indicating significant improvement of the fitting (see Table~\ref{tab-chi2} and Fig.\ref{fig-chi2} GM1). Then we also simulated the eclipse light curves and compared them to the observation, as shown by the blue solid lines in Fig.\ref{fig-eclips-m10}. Now because of the rapid response of Layer-3, the simulated eclipse light curves are much more consistent with the observation. This is also confirmed by the $\chi^2$ improvement of 1037.1 and 48.3 for the eclipse light curves in EPIC-pn and ACIS, separately. Therefore, it is evident that the halo variability due to the eclipse prefers the existence of Layer-3 near Earth.

Then we took another step forward by adding a fourth layer to the scenario and reran the halo fitting. The results are shown in Fig.\ref{fig-mlayers},\ref{fig-chi2} and Table~\ref{tab-bestfit-m10},\ref{tab-chi2}. In this 4-layer best-fit model, Layer-1 is still local to \axj\ with 40.4\% LOS dust. It is also necessary to have a layer local to Earth (i.e. Layer-4) but with a slightly lower dust fraction of 26.9\%. Layer-2 and Layer-3 are on the half-way. Layer-2 is located at a fractional distance of 0.840 and is geometrically thin, containing only 9.0\% of the LOS dust. Layer-3 lies within a fractional distance of 0.104-0.384 and contains 23.7\% LOS dust. The Bayesian analysis reported similar results (Table~\ref{tab-chi2}). But the marginalized distributions of layer boundaries are much wider than in the 2-layer and 3-layer scenarios, indicating much more severe model degeneracies (Fig.\ref{app-fig-mcmc-4p}). However, the red lines in Fig.\ref{fig-eclips-m10} show that the simulated eclipse light curves of the best-fit 4-layer model are not very different from the 3-layer model. Indeed, the $\chi^2$ only decreases by 8.2 (2.0 $\sigma$ for 3 extra d.o.f) and 4.5 (1.2 $\sigma$) for the non-eclipse and eclipse radial profiles, indicating no significant improvement. There is no obvious improvement in the $\chi^2$ of eclipse light curves, either.Therefore, adding more layers does not bring further major improvement, instead the parameter degeneracy will increase rapidly, so we did not add any other additional dust layers.

\subsubsection{Halo Models with Other Types of Dust Grain Populations}
\label{sec-fit-others}
In order to understand the effect of changing the grain population, and to check if the identification of the dust layer near Earth is still robust for different grain configurations, we tried a limited number of cases. We started from the basic MRN dust grain population, assuming that all the layers contain the same grain population (hereafter: GM2 grain configuration, see Table~\ref{tab-modelnames}). We also tried 2-layer, 3-layer and 4-layer scenarios one by one. The best-fit halo models are similar to the above GM1 case. As shown in Fig.\ref{fig-chi2}, we also found a major improvement of the modelling by changing from the 2-layer scenario to 3-layer, with one layer lying within a fractional distance of 0.041 and containing 27\% of the LOS dust, but we did not find obvious improvement by adding more layers. Compared to the GM1 configuration, GM2 configuration produces worse match to the eclipse light curves, as shown in Fig.\ref{app-fig-eclips-GM2}.

Then we switched back to the GM1 configuration, but changed the dust grain population to COMP-NC-B for all layers except Layer-1 (i.e. the one local to the source; hereafter: GM3 grain configuration, see Table~\ref{tab-modelnames}). This is motivated by the recent work of Hankins et al. (2017) where they reported larger typical dust grain size in the Galactic disk than in the GC, and the COMP-NC-B grain population contains the highest fraction of large grains among all the ZDA04 grain populations. For GM3, we also found a major improvement of the modelling for the 3-layer scenario, as shown in Fig.\ref{fig-chi2} and \ref{app-fig-eclips-GM3}, with one layer lying within a fractional distance of 0.047 and containing 44.3\% of the LOS dust. But the best-fit 4-layer model does not bring further improvement to the modelling. As a further test, we continued to change the grain population of Layer-1 to the BARE-GR-B grain population (hereafter: GM4, see Table~\ref{tab-modelnames}), which contains the highest fraction of small dust grains among all the ZDA04 grain populations; while the other layers still used COMP-NC-B. We also found the best-fit 3-layer and 4-layer models are clearly better than 2-layer (Fig.\ref{app-fig-eclips-GM4}), and there is always a layer within a fractional distance of $<~0.1$ to Earth (Table~\ref{app-tab-bestfit-others}). However, neither GM3 nor GM4 is better than GM1 in terms of fitting the halo profiles and eclipse light curves.

In conclusion, changing the dust grain population for different layers can affect the halo fitting, but our tests consistently find that the 3-layer and 4-layer scenarios are clearly better than 2-layer. The improvement is mainly attribute to the inclusion of the layer local to Earth, lying within a fractional distance of $<~0.1$. Adding more layers to the scenario does not bring further improvement.

\section{Discussion}
\label{sec-discussion}
\subsection{Systematic Uncertainties and Model Limitations}
\label{sec-uncertainty}
There are various potential systematic uncertainties associated with our modelling results, one of them originates from the dust grain population. For instance, J17 tried 19 popular dust grain populations and found a large scatter in the best-fit values, except for the robust detection of two major layers and their relative dust fraction. With more than one dust layer, it is possible to use different dust grain populations for different layers. Indeed, it is known that the metallicity close to the GC is higher than Solar values (Anders \& Grevesse 1989; Kubryk, Prantzos \& Athanassoula 2015). The average dust grain size in the GC may also be smaller than typical values in the Galactic disk, which is probably due to the strong turbulence in the CMZ, which can destroy large grains via grain-grain collisions (Hankins et al. 2017). Thus a single grain population must be too simple for the entire GC LOS. However, there are a huge number of different combinations of dust grain populations for multiple layers\footnote{For instance, there will be 20$^{n}$ different combinations for 20 dust grain populations and $n$ layers.}. Different combinations mainly cause scatter in the relative dust fraction in different dust-layers, as well as changing the parameters of the two intermediate dust layers in the four-dust-layer scenario (see Table~\ref{app-tab-bestfit-others}). Since it is unrealistic to try all the combinations due to the computational complexity, this systematic uncertain is difficult to assess. Future investigations about the dust grains in the GC LOS may add more independent constraints.

Additionally, the existence of a halo wing component, whose origin is still unknown, can also introduce systematic uncertainties. We do not know how much LOS dust is responsible for this halo component, but it can certainly reduce the percentage of dust in the other dust layers. There are other systematic uncertainties, such as the assumption about the smooth dust distribution within each geometrically thick dust layer, the approximation of using an effective energy and a specific time point to model the halo variability. All these uncertainties might be partly responsible for the residuals seen between the predicted and observed eclipse light curves in Fig.\ref{fig-eclips-m10} and \ref{app-fig-eclips-GM2}-\ref{app-fig-eclips-GM4}, which we cannot remove.

One obvious limitation of our modelling is that we only tried four model configurations, so we cannot discriminate the type of dust grain population for different layers. However, since it is known that different dust grain populations can produce broadly similar halo components (see Figure 5 in J17), we expect the detection of the three major layers to be robust. Another limitation is related to the remaining significant residuals in the best fit. This is caused by all the higher-order complexities in the data. These remaining complexities can possibly be explained by e.g. adjusting the dust grain models, testing more grain combinations for multiple layers, using non-smooth grain distribution for individual layers, etc. The model approximations adopted here to reduce the computing time (e.g. PSF, energy, time) may also explain some of these higher-order residuals. But addressing these complexities are beyond the objective of this work.

\subsection{Constraining the Distribution of Interstellar Dust towards the GC Direction}
\label{sec-gcdust}
Observing interstellar dust and determining its location is a difficult task. But dust grains are typically contained in the molecular clouds (hereafter: MCs), so the distribution of MCs can be used to infer the distribution of dust. Previously, the existence of the CMZ often led to the speculation that most of the dust along the GC LOS should be located in the inner few hundred parsecs close to \sgra\ (Tan \& Draine 2004). However, recent studies reported different results. Firstly, many studies about the radio emission from the GC magnetar SGR J1745-2900, (2.4 arcsec away from \sgra) showed that the majority of scattering gas should be located in the Galactic disk far from the GC (Kennea et al. 2013; Mori et al. 2013; Rea et al. 2013; Eatough et al. 2013; Shannon \& Johnston 2013; Bower et al. 2014; Wucknitz 2015; Sicheneder \& Dexter 2017; Dexter et al. 2017; but see Spitler et al. 2014). Since gas and dust should be physically connected in the ISM (e.g. Predehl \& Schmitt 1995; Draine \& Bone 2004; G\"{u}ver \& \"{O}zel 2009; Valencic \& Smith 2015; Zhu et al. 2017), it can be speculated that most of the dust should also reside in the Galactic disk. Secondly, studies about the GC IR extinction also suggested that most of the dust extinction should occur in the Galactic disk (Fritz et al. 2011; Voshchinnikov, Henning \& Il'in 2017). Most recently, J17 modelled the X-ray dust scattering halo around \axj, and provided strong evidence that (66-81)\% of the GC LOS dust should be located in the Galactic disk rather than in the GC$^{\small\ref{fn-dust-frac}}$.

Indeed, previous studies using molecular lines to trace the spatial distribution of MCs revealed a massive molecular ring structure at 5 kpc away from the GC, which seems to be distributed in the Galactic disk along the Scutum Arm (Jackson et al. 2006; Simon et al. 2006; Roman-Duval et al. 2010; Dobbs \& Burkert 2012; Sato et al. 2014; Heyer \& Dame 2015). Surveys about the distribution of IR dark clouds in the Milky Way also find that they link to the Scutum Arm (Marshall, Joncas \& Jones 2009; Peretto \& Fuller 2010). Therefore, there should be a large amount of MCs in the Galactic disk where the interstellar dust can be contained.

\begin{figure}
\centering
\includegraphics[bb=75 0 612 630, scale=0.35]{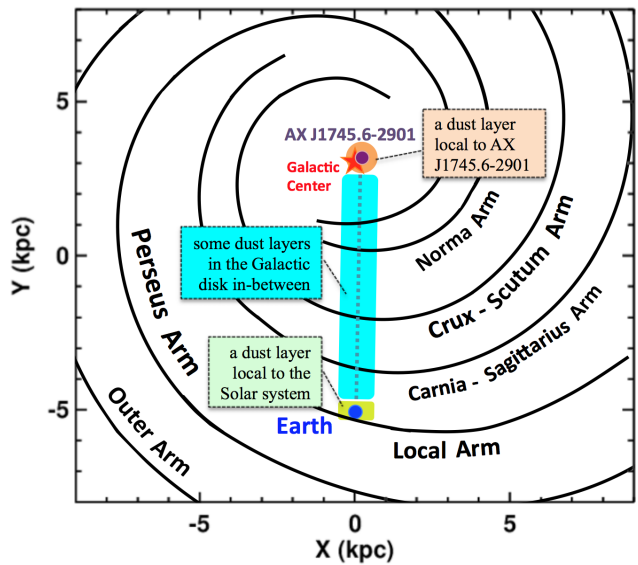}
\caption{Dust distribution along the LOS of \axj, as suggested by the best-fit 3-layer and 4-layer models in Table~\ref{tab-bestfit-m10}, assuming it is in the GC. The schematic of Milky Way's spiral arms, as shown in Nakanishi \& Sofue (2016), Reid et al. (2009), Caswell \& Haynes (1987) (also see the review by Bland-Hawthorn \& Gerahrd (2016)), are also plotted for comparison. Compared to the 2-layer model in Figure 15 of J17, now we identified a third dust layer in the Solar neighbourhood (see Section~\ref{sec-dsn}), although the second dust layer (cyan region) in the Galactic disc is still poorly resolved.}
\label{fig-gcstruct}
\end{figure}

The main objective of this work is to use the dust scattering model to explain the peculiar eclipse phenomena of \axj. As a side produce, this also allows us to obtain further constraints on the LOS dust distribution. We confirm that the 2-layer models in J17 do reflect the general LOS dust distribution towards \axj, and we can further resolve the foreground dust layer into two separate layers, as shown in Fig.\ref{fig-gcstruct}. Our best-fit halo models indicate that one layer is local to Earth and contains (20-50)\% of the total LOS dust, which is likely to be associated with the MCs in the Solar neighbourhood (see next section). Another (20-40)\% dust is local to \axj, and the remaining dust should reside in the Galactic disk in-between. The best-fit 4-layer models indicate that one intermediate dust layer resides within a fractional distance of 0.1-0.5, while the other intermediate layer lies within 0.5-0.9, but we cannot obtain further constraints due to severe model degeneracies, so the ISM in any of the intervening spiral arms can be responsible for the intermediate dust layer, such as the near 3-kpc arm, Normal Arm, Scutum Arm and Carnia-Sagittarius Arm (see Fig.\ref{fig-gcstruct}). While the detection of these major layers appear robust, the fraction of dust contained in each layer shows large scatter between different best-fit models. This scatter is due to various systematic uncertainties which are discussed in Section~\ref{sec-uncertainty}.

Those GC foreground dust layers far from \axj\ should also intervene in the LOS to some other sources with small angular separations to \axj. If we adopt the X-ray absorbing hydrogen column density\footnote{$N_{\rm H, abs}$ values are all based on the {\sc tbnew} model with cross-sections of Verner et al. (1996) and abundances of Wilms, Allen \& McCray (2000).} ($N_{\rm H, abs}$) of \axj\ to be $\sim3\times10^{23}$ cm$^{-2}$ (J17), then all the dust layers far from \axj\ would correspond to $N_{\rm H, abs}=(1.8-2.4)\times10^{23}$ cm$^{-2}$. Indeed, it has been reported that \sgra\ has $N_{\rm H, abs}\sim1.6\times10^{23}{\rm cm}^{-2}$ (Ponti et al. 2017b), the GC LMXB Swift J174540.7-290015 at 16.6 arcsec away from \sgra\ has $N_{\rm H, abs}\sim1.8\times10^{23}{\rm cm}^{-2}$ (Ponti et al. 2016), and the magnetar SGR J1745-2900 at 2.4 arcsec away from \sgra\ also has $N_{\rm H, abs}\sim1.8\times10^{23}{\rm cm}^{-2}$ (Coti Zelati et al. 2015; Ponti et al. 2017b). These sources all have similar $N_{\rm H, abs}$ to that contained in the GC foreground dust layers, which is consistent with the possibility that their LOSs all go through these foreground layers.

\subsection{The Dust Layer in the Solar Neighbourhood}
\label{sec-dsn}
The MC distribution in the Milky Way has been mapped in great detail by many surveys during the past few decades, using either molecular line emission (e.g. $^{12}$CO, $^{13}$CO, HCO$^{+}$, Jones et al. 2012; also see the review by Heyer \& Dame 2015), or interstellar dust emission (e.g. Schlegel et al. 1998; Planck Collaboration et al. 2016), or dust reddening (e.g. Lallement et al. 2014; Alves et al. 2014; Green et al. 2015). Since the Solar system is only $\sim$25 pc above the Galactic plane (Goodman et al. 2014), and the thickness of the molecular layer in the Galactic disk is $\lesssim$ 120 pc for the Galactic radii of $\lesssim$ 8 kpc (Heyer \& Dame 2015), the LOS from Earth to the GC should intersect with the MCs in the Galactic disk. It has also been found that our Solar system resides in a local cavity of ISM of 80-200 pc wide (i.e. the `Local Bubble', e.g. Tanaka \& Bleeker 1977; Snowden et al. 1990; Bergh\"{o}fer \& Breitschwerdt 2002). Outside this cavity there is a luminous elliptical ring structure containing many young stars (Gould Belt) and dense ISM (Lindblad Ring) (e.g. Poppel 1997; Perrot \& Grenier 2003; Lallement et al. 2014; Chen et al. 2016; Broekhoven-Fiene et al. 2018). Therefore, despite the large scatter and systematic uncertainties in our best-fit parameters (see next section), we can check if the nearby dust layer can be associated with any known nearby MCs.

Our halo analysis found that the mid-point of the nearby dust layer is at a fractional distance of $\sim$ 0.02. Assuming \axj\ is 8 kpc away, this mid-point would be at $\sim$ 160 pc from Earth, which is roughly consistent with the distance to the Ophiuchus MC (140-150 pc, Ortiz-Le\'{o}n et al. 2017) in the Gould belt in the direction of longitude $l\sim0\degr$. However, the Ophiuchus MC is at the latitude $b\sim10-20\degr$, so it needs to be $\gtrsim$ 80 pc wide in order to extend into the LOS of \axj. Miville-Desch{\^e}nes, Murray \& Lee (2017) studied 8107 MCs in the Galactic plane, and found that the cloud radius ranges from less than 1 pc to a few hundred pc, with an average value of 30 pc, and thus a 80 pc MC is not rare. Another possibility is that this dust layer is associated with the MCs in the Local Arm, but not necessarily in the Gould belt. The three-dimensional map of the local ISM produced by Lallement et al. (2014) reveals that in the GC direction ($l\sim0\degr$, $b\sim0\degr$), the differential dust reddening opacity strongly enhances within the region of 100-400 pc from Sun, with at least two major peaks at $\sim$130 pc and $\sim$ 220 pc (see their Figure 4). Thus there are indeed MCs along the LOS of \axj\ within 400 pc to account for the dust layer in the Solar neighbourhood. However, considering systematic uncertainties in our modelling, together with the difficulty of estimating the dust column in different MCs distributed along the GC LOS, we cannot provide stronger evidence for the connection between this best-fit nearby dust layer and nearby MCs.

Furthermore, one important inference of this dust layer in the Solar neighbourhood is that it may cover a relatively large sky area in the GC direction. Assuming that the MC is at 160 pc away and has a radius of 1-100 pc, it will cover an circular region of 0.4-32 degree radius around \axj, which includes \sgra\ and many other GC sources. Then the dust scattering halo study for a sample of GC sources can be used to verify the existence of this nearby dust layer, and possibly constrain the size of the corresponding MC and the fraction of LOS dust and gas contained in it. This is beyond the scope of this work and will be reported in our next paper.

\subsection{Effects of Dust Scattering on the Measurement of Eclipse Ingress and Egress Times}
\label{sec-egress}
Our study has fully demonstrated the dust scattering origin for the peculiar shape of the eclipse light curves of \axj, but another important phenomenon about the orbital period evolution of \axj\ is not understood yet. Ponti et al. (2017a) used the Bayesian Block method to single out the eclipse ingress and egress times, which led to the discovery of a significant long-term decreasing rate of its orbital period and 10-20 s jitter. From our simulations in Fig.~\ref{fig-halovar}, it is clear that the dust scattering halo will show a lagged response to the eclipse signal, which may affect the determination of eclipse ingress and egress times. However, Ponti et al. (2017a) mainly used \xmm\ data and extracted 3-10 keV source light curves from a circular region with 40 arcsec, where the primary source emission dominated over the halo emission{\footnote{Ponti et al. (2017a) also adopted 1 {\it ASCA} observation, from which they extracted 3-10 keV source light curves from a circular region with 0.8 arcmin radius where the primary source emission also dominated.}}. In this case, the instantaneous flux drop and rise, as caused by the primary source eclipse, are conspicuous and can be easily identified by the Bayesian Block method as the eclipse ingress and egress times. The presence of dust scattering halo only changed the shape of the eclipse edge from a right angle to a curved angle, as shown in their Figure 1, but did not change the eclipse ingress and egress times. Besides, the effects of dust scattering should be the same for every orbit. Therefore, we can safely rule out dust scattering as the cause of the puzzling orbital period decrease and jitter of \axj.

Nonetheless, our study raises the importance of considering the timing effects of dust scattering halo when studying the variability of other X-ray sources in the GC, especially \sgra\ whose LOS is close to \axj. For instance, the profile of X-ray flares from \sgra\ must also be affected by the timing effects of the dust scattering halo created by multiple foreground dust layers, but this requires a separate study in the future.

\section{Summary and Conclusions}
In this paper we used \xmm\ and \cxo\ observations and the X-ray dust scattering theory to perform a thorough analysis of the peculiar eclipse phenomena of \axj. The main results of our study are summarised below.
\begin{itemize}
\item Based on \xmm\ and \cxo\ observations, we confirmed previous reports about the significant excess flux during the eclipse phase of \axj. Moreover, our detailed analysis showed that the eclipse flux and shape of the eclipse light curves of \axj\ strongly depend on both photon energy and the shape of the source extraction region. The instrumental PSF will also affect the shape of the observed dust scattering halo, so that different instruments will also observe different shapes of eclipse light curves.
\item By performing detailed simulations for the eclipse signal of \axj, we showed that the dust scattering halo shows different variability for dust layers at different locations along the LOS. When the dust layer is not local to the source, the eclipse signal can be observed to {\it propagate} from inner to outer radii of the halo due to the increasing time lag, similar to the expanding dust scattering rings observed around some fast X-ray transients. The simulated halo profile and variability also depend on the photon energy and instrumental PSF.
\item We performed simultaneous modelling for 21 different halo profiles of \axj\ observed by \xmm\ and \cxo\ in the 2-4, 4-6 and 6-10 keV bands during non-eclipse and eclipse phases. With three to four dust layers, our modelling reproduced the energy, instrument and radial dependences of the observed eclipse light curves, confirming the dust scattering origin of the peculiar eclipse phenomena of \axj. The characteristic eclipse light curves reported in this paper can also be used to verify the origin of dust scattering in other eclipsing sources, provided that there is significant amount of dust along the LOS.
\item The modelling of the halo variability also provides further constraints on the LOS dust distribution of \axj. We confirmed a dust layer local to \axj\ and containing (20-40)\% of the LOS dust as first reported by J17. We also identified another separate dust layer being only a few hundred parsecs from Earth and containing up to a few tens of percent LOS dust. The remaining several tens of per cent LOS dust is located in-between in the Galactic disk. The existence and separation of these major dust layers appear robust, but their dust fractions are subject to various potential systematic uncertainties, which we also discussed in detail. Moreover, we discussed the possibility that the nearby dust layer can be associated with some MCs in the Solar neighbourhood, probably located in the Local Arm and/or the Gould Belt/Lindblad Ring.
\item We explained that the dust scattering halo does not affect the accuracy of the measurements of eclipse ingress and egress times in \axj, and so it cannot be the cause for the puzzling orbital period decrease and jitter of \axj\ as reported by Ponti et al. (2017a). But our study raises the importance of considering the timing effects of dust scattering for the variability study of other X-ray sources in the GC, such as the study of X-ray flares from \sgra.
\end{itemize}

The exploration of dust scattering halos and dust grain properties in the GC direction just begins and requires lots of future work. In Section~\ref{sec-discussion} we pointed out the importance of studying the halo profile for a sample of bright GC X-ray sources, which will be reported in our next paper. Moreover, in order to better resolve the dust distribution along the GC LOS, it would be highly valuable to observe the dust scattering around very bright GC X-ray sources with very strong variability, most ideally being some energetic and fast X-ray transients. The strong X-ray pulsed signals from such sources can create dust scattering ring structures, which would allow robust identifications and precise distance measurements for those multiple GC foreground dust layers, as well as the absolute source distances. Since new activities of GC transients emerge almost every year, such extreme source variabilities may be observed in the future.

\section*{Acknowledgements}
We thank our anonymous referee for providing valuable comments and suggestions. CJ acknowledges inspiring discussions with Guangxing Li about the molecular clouds in the Milky Way. CJ and GP acknowledge Kishalay De for helpful discussions about the eclipse timing. This project is supported by the Bundesministerium f\"{u}r Wirtschaft und Technologie/Deutsches Zentrum f\"{u}r Luft- und Raumfahrt (BMWI/DLR, FKZ 50 OR 1408 and FKZ 50 OR 1604) and the Max Planck Society. This work is based on observations obtained with {\it XMM-Newton}, an ESA science mission with instruments and contributions directly funded by ESA Member States and NASA. The scientific results reported in this article are also based on observations made by the Chandra X-ray Observatory, as well as data obtained from the Chandra Data Archive. This research has made use of software provided by the Chandra X-ray Center (CXC) in the application packages CIAO, ChIPS, and Sherpa.




\appendix

\section{Simultaneous Fitting to the Eclipsing Radial Profiles}
\begin{figure*}
\begin{tabular}{c}
\includegraphics[bb=54 36 486 570, scale=0.9]{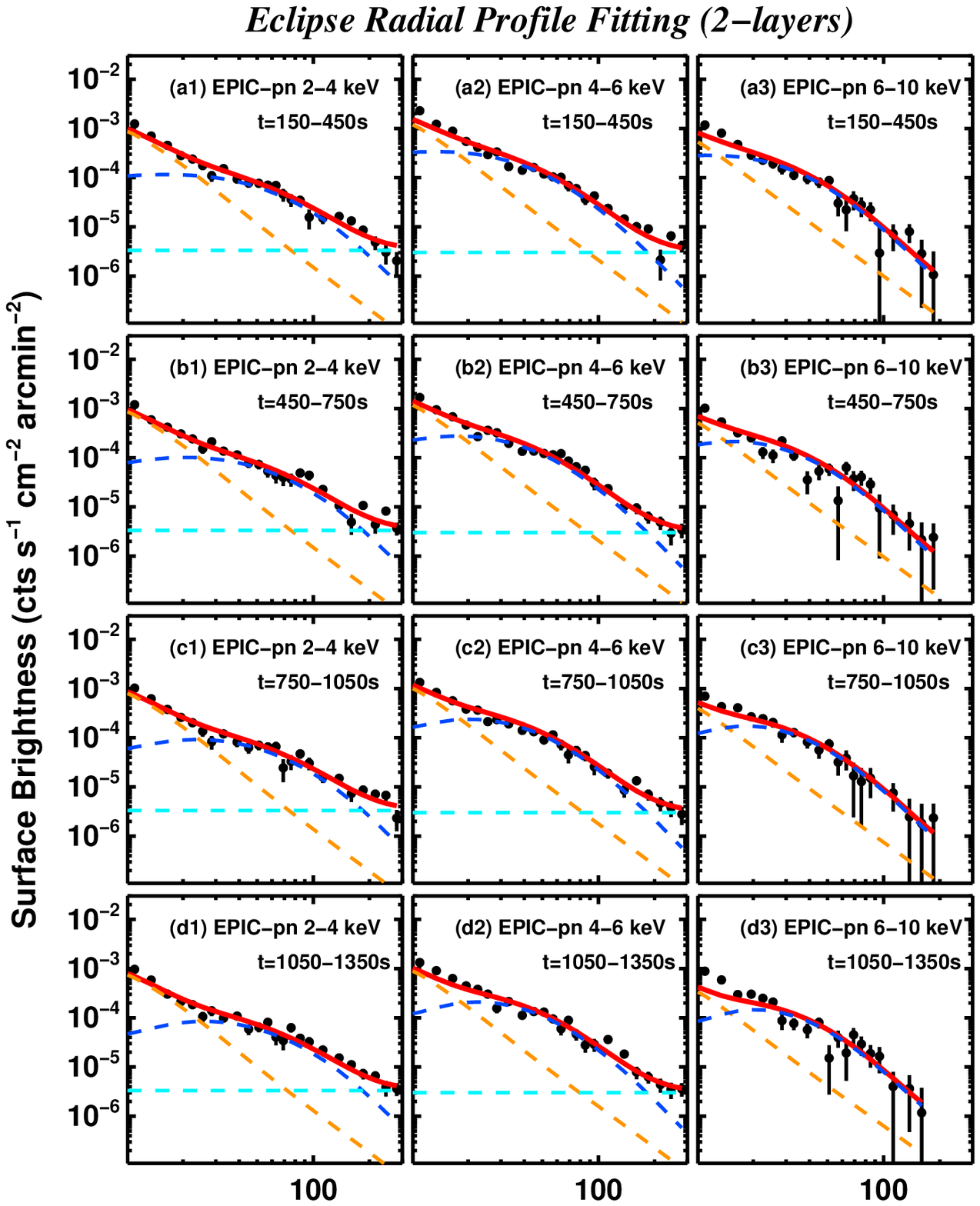} \\
\includegraphics[bb=54 36 486 163, scale=0.9]{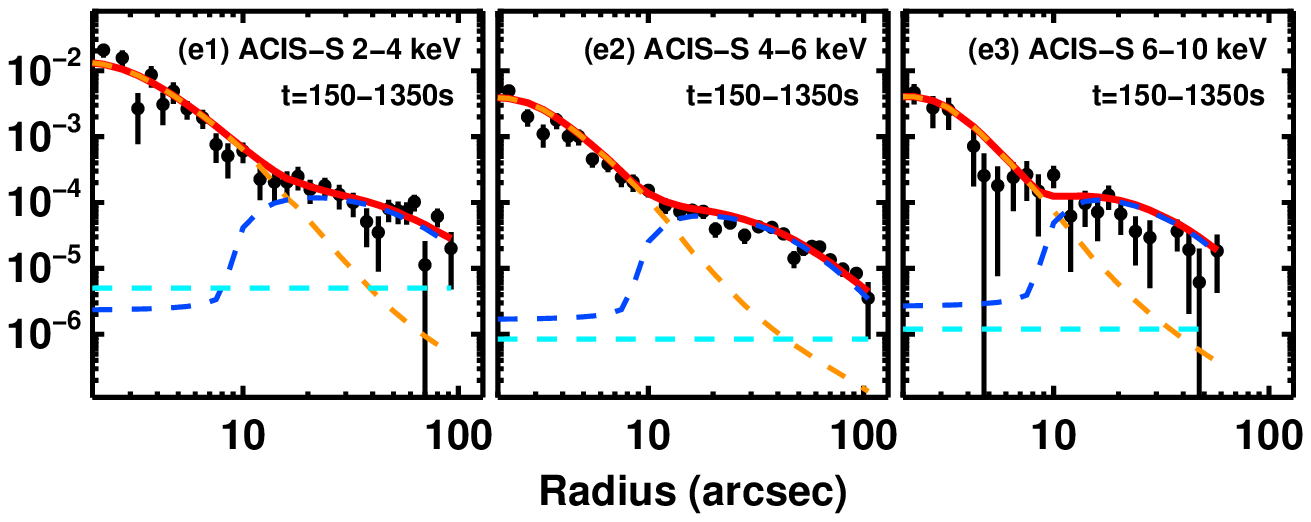} \\
\end{tabular}
\caption{A simultaneous fit to the non-eclipse radial profiles in Fig.\ref{fig-mlayers} and eclipse radial profiles in this figure, using the 2-layer scenario with the COMP-AC-S grain population in every layer. The type of instrument, energy band and time interval (relative to the eclipse ingress time) have all been labeled in every panel. The orange dash line is for Layer-1, blue is for Layer-2, and cyan is for the halo wing component. The red solid line is for the total halo model. There is no PSF component in the radial profile as the primary source is eclipsed. The best-fit parameters can be found in Tables~\ref{tab-bestfit-m10} and \ref{tab-chi2}.}
\label{app-fig-fit-eclips-2p}
\end{figure*}

\begin{figure*}
\begin{tabular}{c}
\includegraphics[bb=54 36 486 570, scale=0.9]{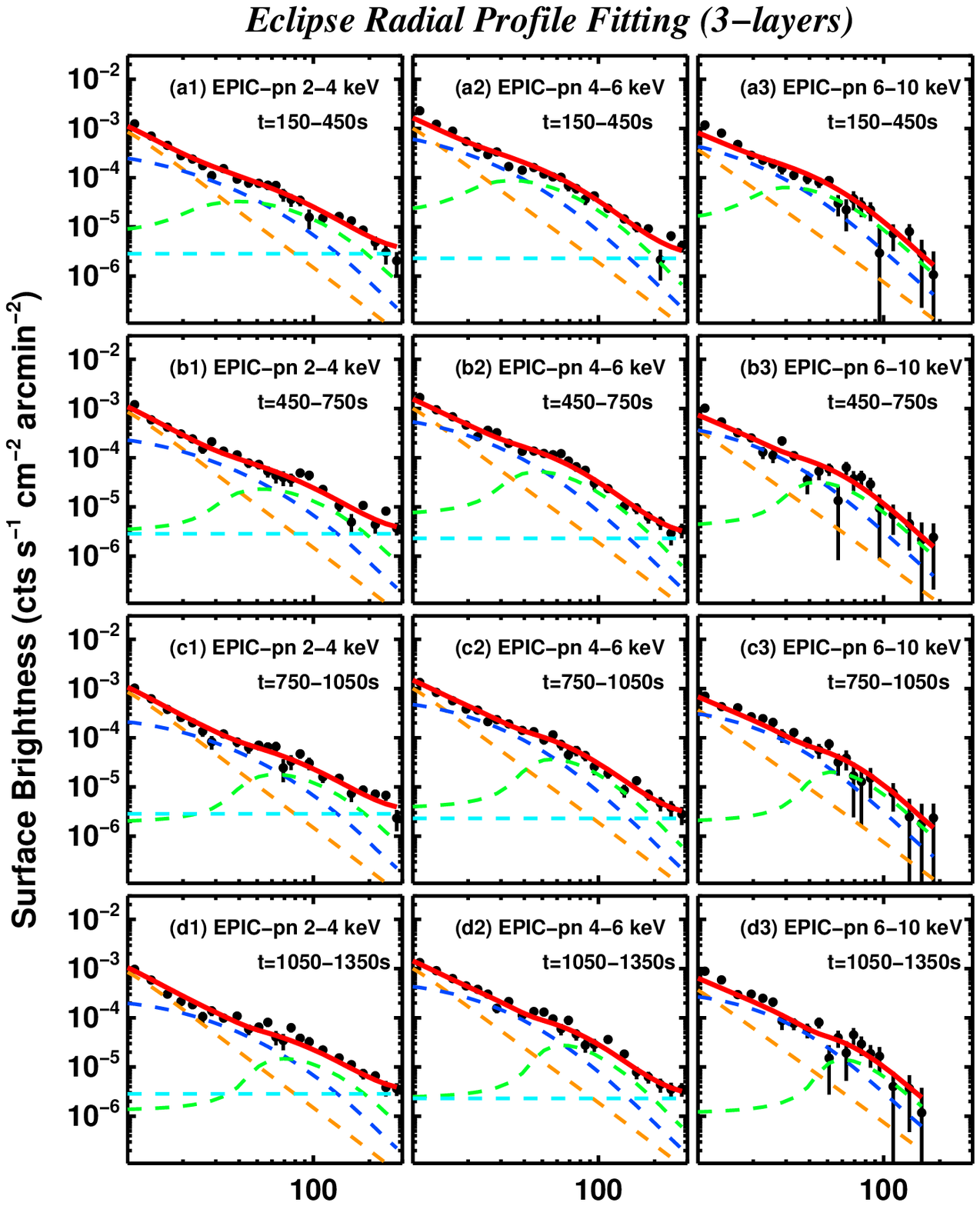} \\
\includegraphics[bb=54 36 486 163, scale=0.9]{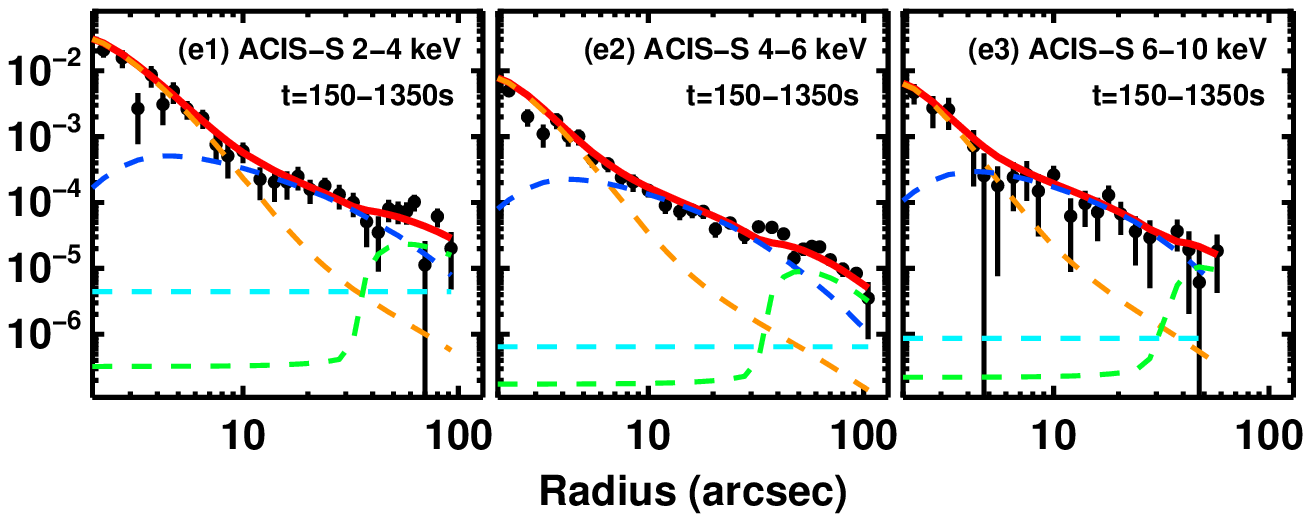} \\
\end{tabular}
\caption{Similar to Fig.\ref{app-fig-fit-eclips-2p}, but with three dust layers along the LOS as shown in Fig.\ref{fig-mlayers}.}
\label{app-fig-fit-eclips-3p}
\end{figure*}

\begin{figure*}
\begin{tabular}{c}
\includegraphics[bb=54 36 486 570, scale=0.9]{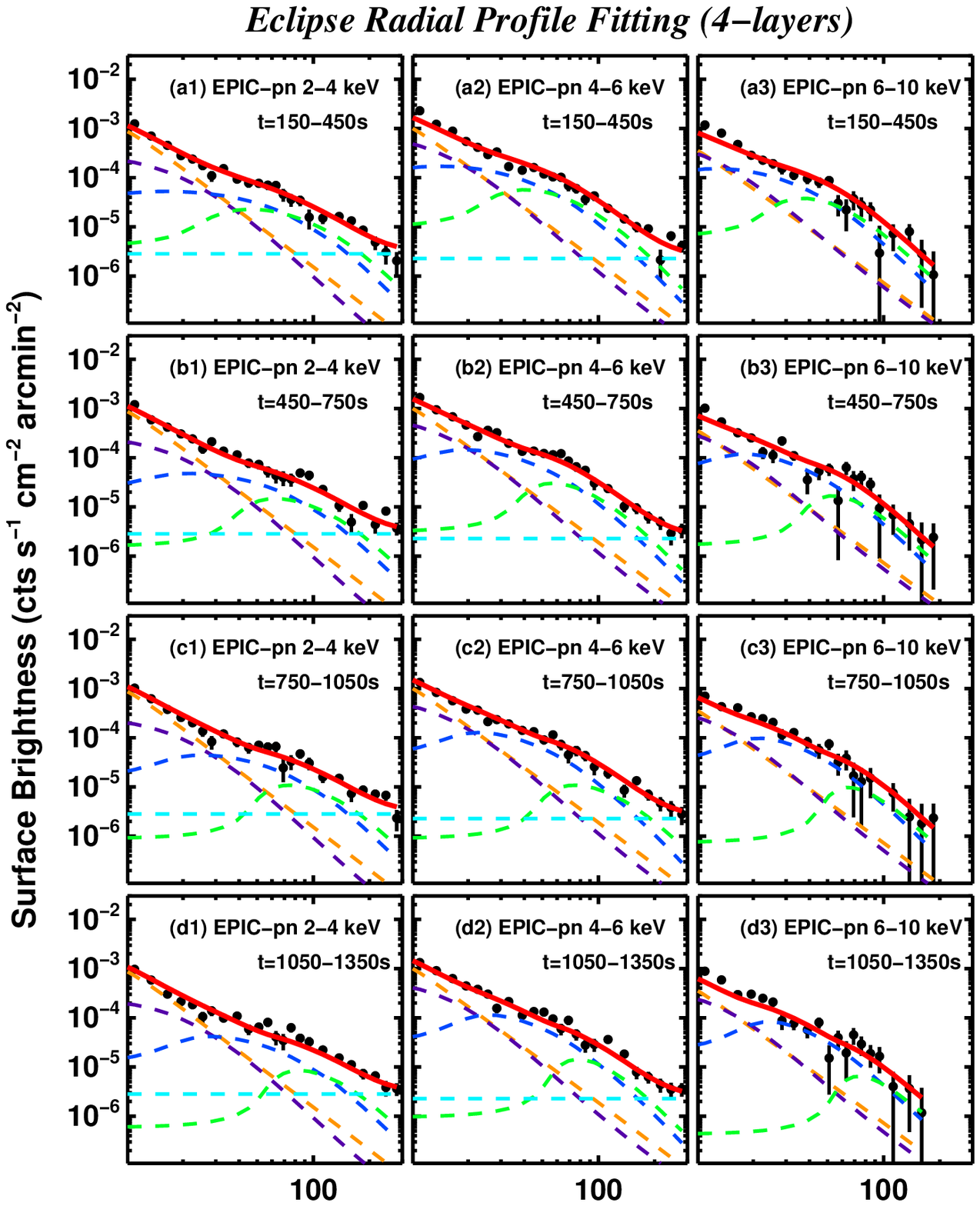} \\
\includegraphics[bb=54 36 486 163, scale=0.9]{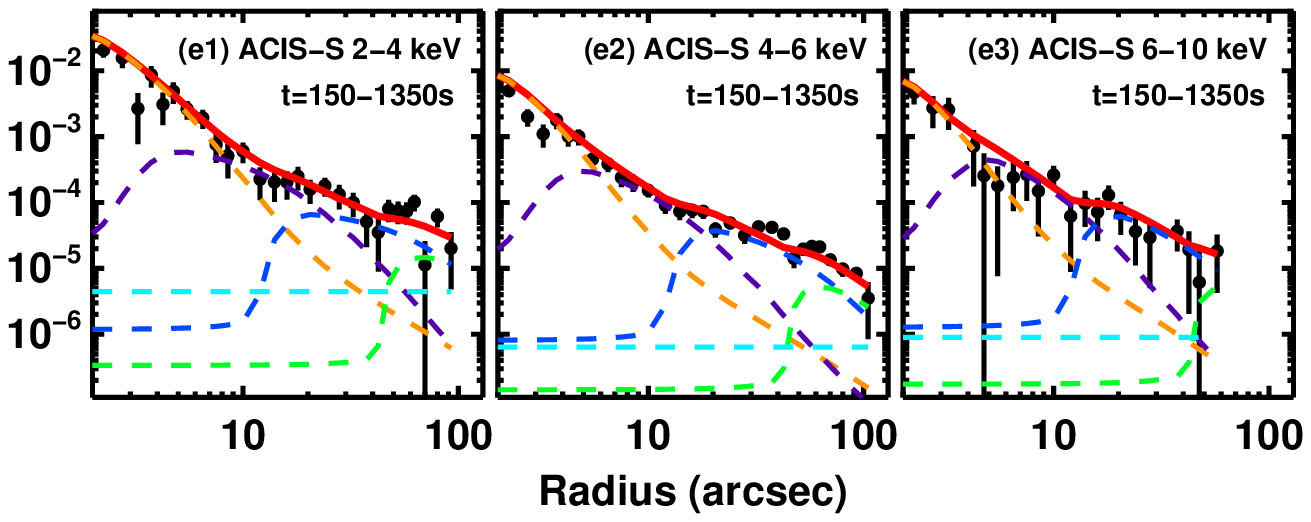} \\
\end{tabular}
\caption{Similar to Fig.\ref{app-fig-fit-eclips-2p}, but with four dust layers along the LOS as shown in Fig.\ref{fig-mlayers}.}
\label{app-fig-fit-eclips-4p}
\end{figure*}

\section{Results of the MCMC Sampling for the COMP-AC-S Dust Grain Model for 3 and 4-layer Scenarios.}
\begin{figure*}
\includegraphics[bb=185 -20 700 930, scale=0.575]{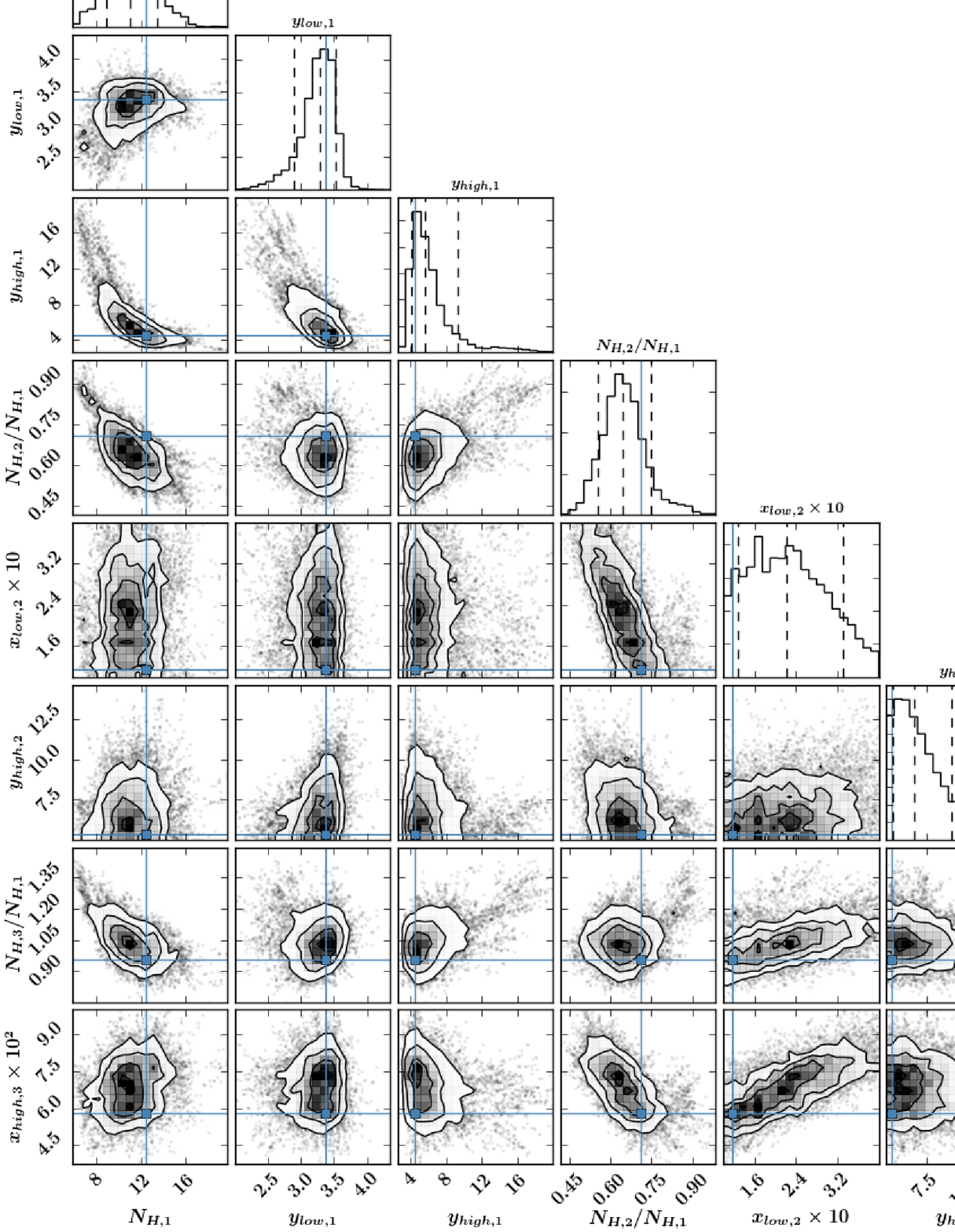} \\
\caption{Similar to Fig.\ref{fig-mcmc-2p}, but for the MCMC sampling under the 3-layer scenario with the COMP-AC-S dust grain model. The meaning of each parameter can be found in Fig.\ref{fig-mlayers}. For clarity, we performed the linear conversion: $y_{\rm low,1}=(1-x_{\rm low,1})\times10^{2}$, $y_{\rm high,1}=(1-x_{\rm high,1})\times10^{3}$, $y_{\rm high,2}=(1-x_{\rm high,2})\times10^{2}$. The blue solid lines indicate the best-fit values using the minimum $\chi^2$ method.}
\label{app-fig-mcmc-3p}
\end{figure*}

\begin{figure*}
\includegraphics[bb=230 -20 700 940, scale=0.54]{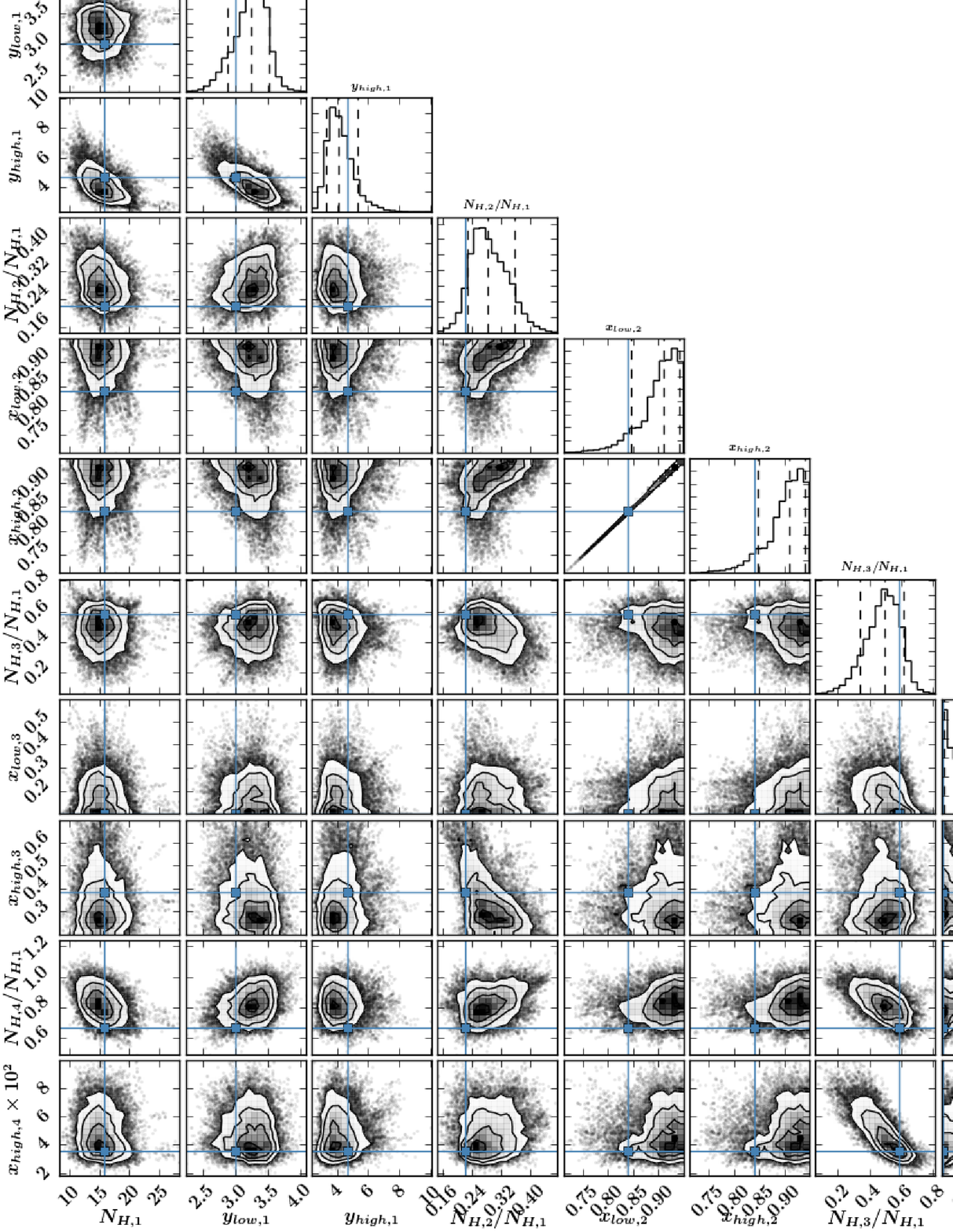} \\
\caption{Similar to Fig.\ref{fig-mcmc-2p}, but for the MCMC sampling under the 4-layer scenario with the COMP-AC-S dust grain model. The meaning of each parameter can be found in Fig.\ref{fig-mlayers}. For clarity, we performed the linear conversion: $y_{\rm low,1}=(1-x_{\rm low,1})\times10^{2}$, $y_{\rm high,1}=(1-x_{\rm high,1})\times10^{3}$. The blue solid lines indicate the best-fit values using the minimum $\chi^2$ method.}
\label{app-fig-mcmc-4p}
\end{figure*}

\section{Comparison between the Observed and Modelled Eclipse Light Curves for Other Dust Grain Models}

\setlength\dashlinedash{0.6pt}
\setlength\dashlinegap{1.5pt}
\setlength\arrayrulewidth{0.8pt}
\begin{table}
\centering
\caption{The best-fit model parameters for the GM2, GM3 and GM4 dust grain configurations (see Section~\ref{sec-results}).}
\begin{tabular}{@{}llcccc@{}}
\hline
& Parameter & GM2 & GM3 & GM4 \\
\hline
\multicolumn{2}{l}{2-layer Scenario} &  \\ 
\hdashline
Layer-1 & x$_{\rm low,1}$ & 0.980 & 0.860 & 0.969 \\
& x$_{\rm high,1}$ & 0.992 & 0.999 & 0.999 \\
& $f_{\rm nH,1}$ (\%) & 33.8 & 30.2 & 37.6 \\
\hdashline
Layer-2 & x$_{\rm low,2}$ & 0-fixed & 0-fixed & 0-fixed \\
& x$_{\rm high,2}$ & 0.914 & 0.158 & 0.302 \\
& $f_{\rm nH,2}$ (\%) & 66.2 & 69.8 & 62.4 \\
\hdashline
& $N_{\rm H,tot}$ ($10^{22}$ cm$^{-2}$) & 14.0 & 48.6 & 54.9 \\
\hline
\multicolumn{2}{l}{3-layer Scenario} & & & \\
\hdashline
Layer-1& x$_{\rm low,1}$ & 0.976 & 0.964 & 0.971 \\
& x$_{\rm high,1}$ & 0.998 & 0.995 & 0.999 \\
& $f_{\rm nH,1}$ (\%)  & 33.7 & 29.3 & 31.5 \\
\hdashline
Layer-2& x$_{\rm low,2}$ & 0.152 & 0.110 & 0.155 \\
& x$_{\rm high,2}$ & 0.948 & 0.901 & 0.845 \\
& $f_{\rm nH,2}$ (\%) & 39.3 & 26.4 & 21.1 \\
\hdashline
Layer-3 & x$_{\rm low,3}$ & 0-fixed & 0-fixed & 0-fixed \\
& x$_{\rm high,3}$ & 0.041 & 0.047 & 0.073 \\
& $f_{\rm nH,3}$ (\%) & 27.0 & 44.3 & 47.4 \\
\hdashline
& $N_{\rm H,tot}$ ($10^{22}$ cm$^{-2}$) & 24.8 & 50.1 & 67.4 \\
\hline
\multicolumn{2}{l}{4-layer Scenario} & & & \\
\hdashline
Layer-1 & x$_{\rm low,1}$ & 0.977 & 0.973 & 0.971 \\
& x$_{\rm high,1}$ & 0.998 & 0.987 & 0.999 \\
& $f_{\rm nH,1}$ (\%) & 33.3 & 24.5 & 31.2 \\
\hdashline
Layer-2 & x$_{\rm low,2}$ & 0.536 & 0.632 & 0.553 \\
& x$_{\rm high,2}$ & 0.947 & 0.944 & 0.824 \\
& $f_{\rm nH,2}$ (\%) & 22.2 & 15.7 & 11.3 \\
\hdashline
Layer-3 & x$_{\rm low,3}$ & 0.152 & 0.116 & 0.156 \\
& x$_{\rm high,3}$ & 0.454 & 0.243 & 0.396 \\
& $f_{\rm nH,3}$ (\%)  & 17.4 & 21.5 & 9.9 \\
\hdashline
Layer-4 & x$_{\rm low,4}$ & 0-fixed & 0-fixed & 0-fixed \\
& x$_{\rm high,4}$ & 0.042 & 0.040 & 0.073 \\
& $f_{\rm nH,4}$ (\%)  & 27.1 & 38.3 & 47.6 \\
\hdashline
& $N_{\rm H,tot}$ ($10^{22}$ cm$^{-2}$) & 23.6 & 35.8 & 69.1 \\
\hline
\end{tabular}
\label{app-tab-bestfit-others}
\end{table}

\begin{figure*}
\begin{tabular}{l}
\includegraphics[bb=-50 216 738 750, scale=0.39]{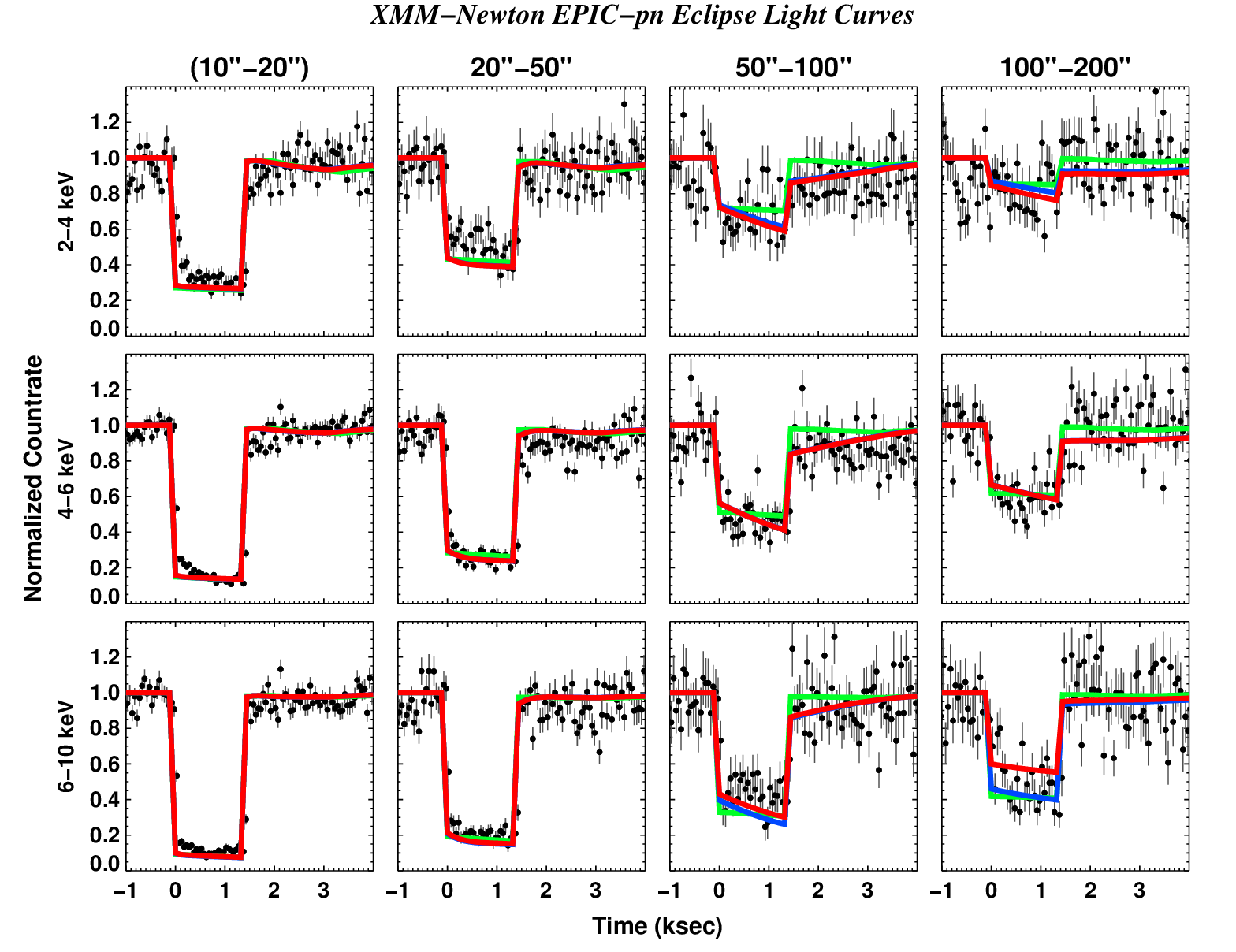} \\
\includegraphics[bb=54 216 954 450, scale=0.39]{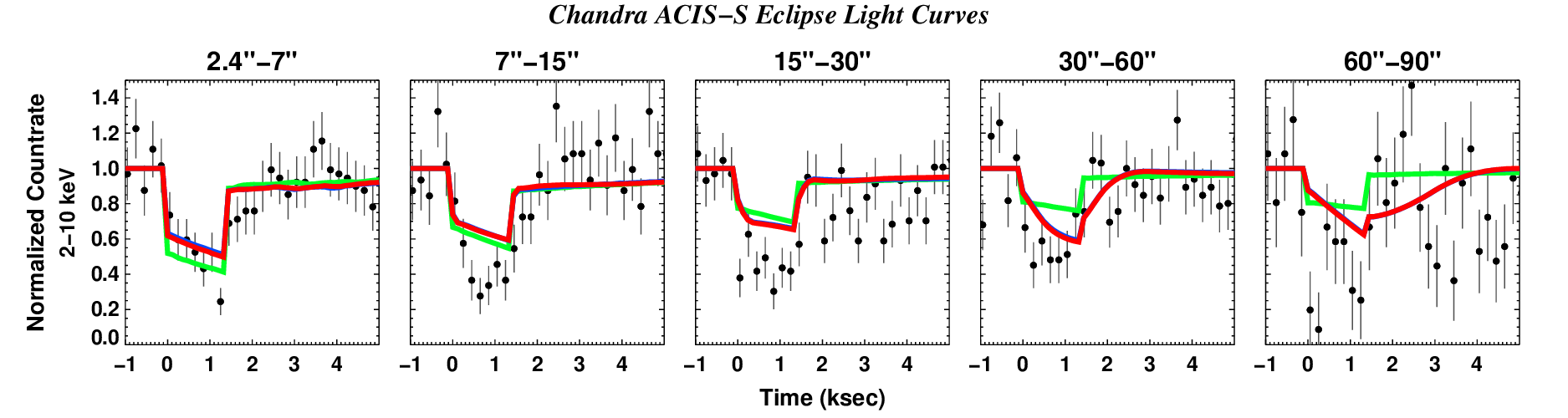} \\
\end{tabular}
\caption{Similar to Fig.\ref{fig-eclips-m10}, but using the GM2 dust grain configuration, i.e. replacing the dust grain model with the MRN grain population for all dust layers.}
\label{app-fig-eclips-GM2}
\end{figure*}

\begin{figure*}
\begin{tabular}{l}
\includegraphics[bb=-50 216 738 750, scale=0.39]{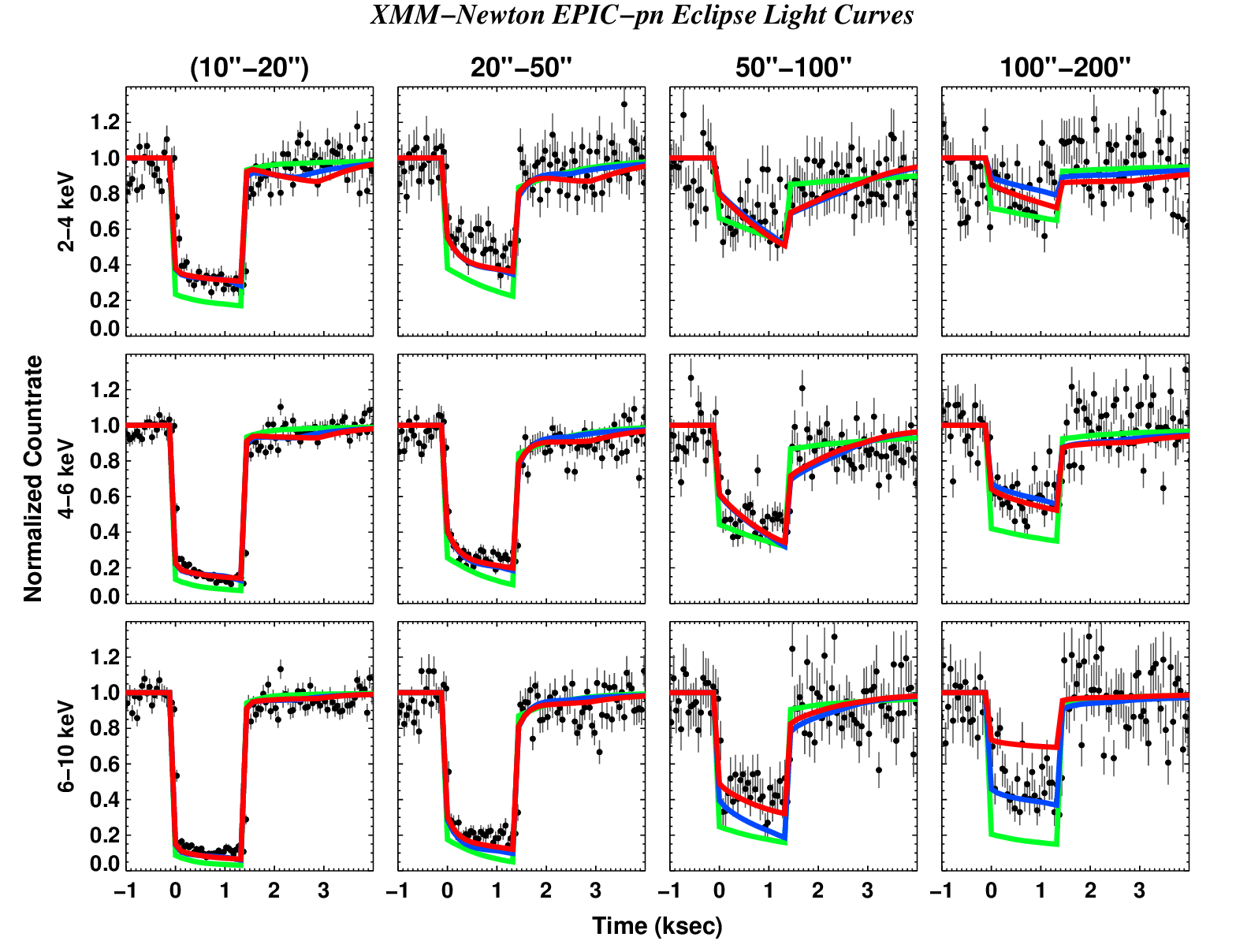} \\
\includegraphics[bb=54 216 954 450, scale=0.39]{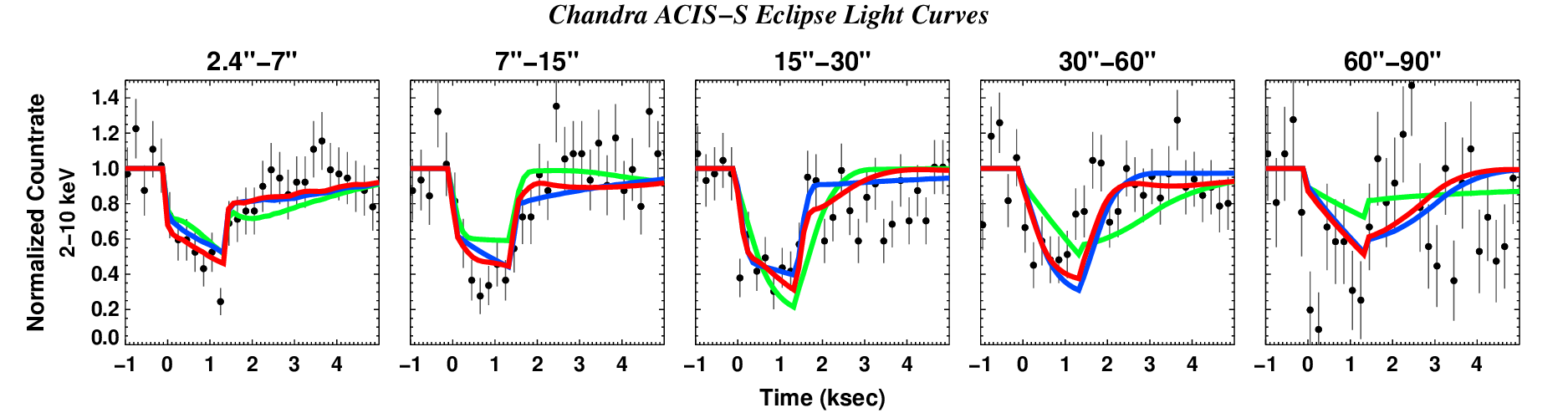} \\
\end{tabular}
\caption{Similar to Fig.\ref{fig-eclips-m10}, but using the GM3 dust grain configuration, i.e. using the replacing the dust grain model with the ZDA04 COMP-NC-B grain population for all the dust layers except the one local to the source.}
\label{app-fig-eclips-GM3}
\end{figure*}

\begin{figure*}
\begin{tabular}{l}
\includegraphics[bb=-50 216 738 750, scale=0.39]{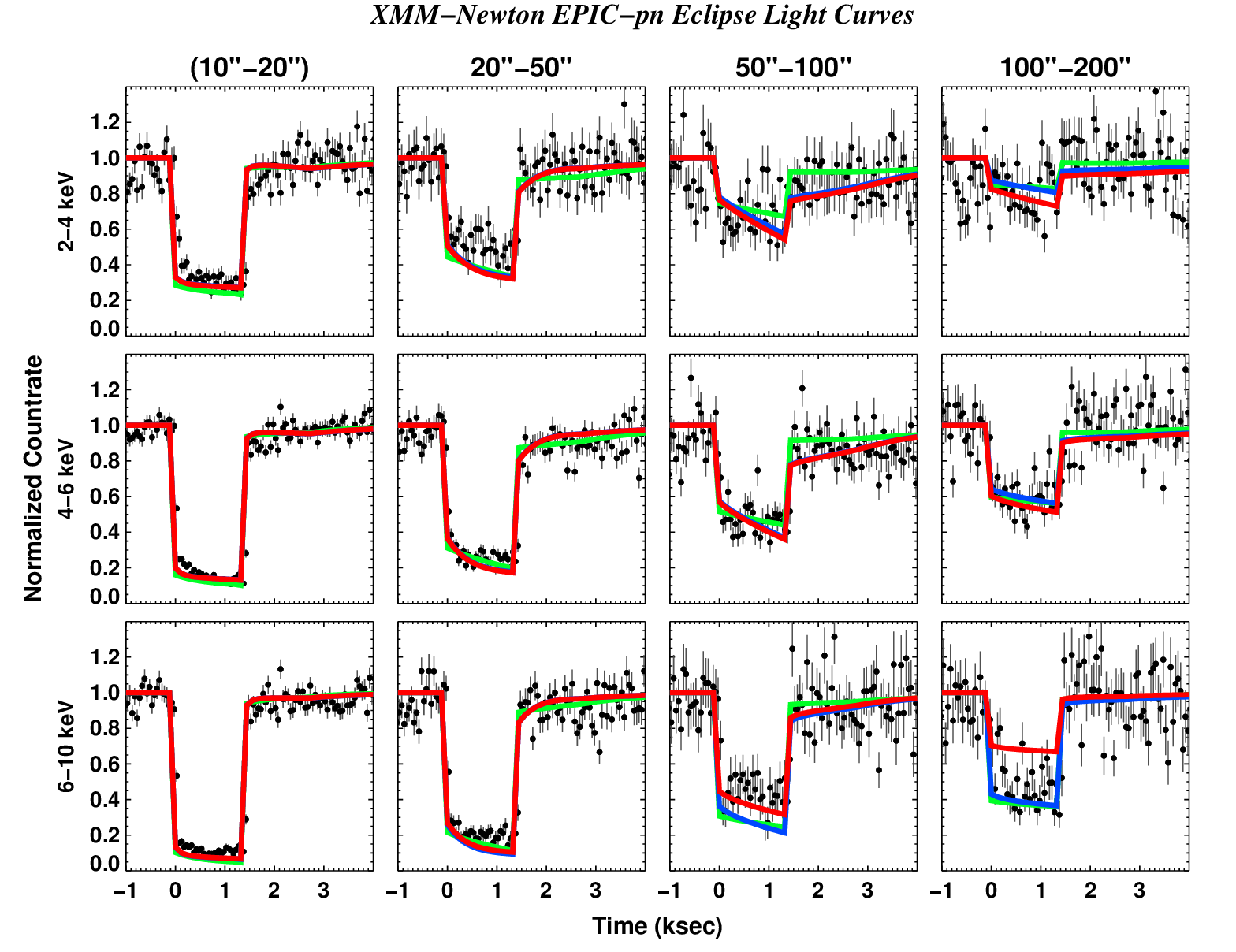} \\
\includegraphics[bb=54 216 954 450, scale=0.39]{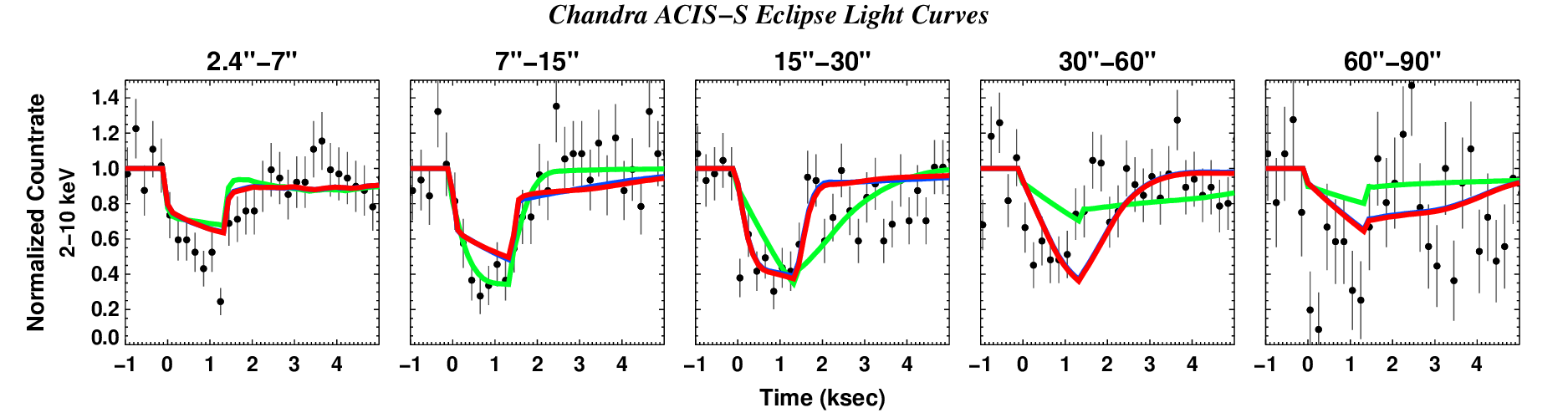} \\
\end{tabular}
\caption{Similar to Fig.\ref{app-fig-eclips-GM3}, but using the GM3 dust grain configuration, i.e. replacing the dust grain model with the ZDA04 BARE-GR-B grain population for the dust layer local to the source.}
\label{app-fig-eclips-GM4}
\end{figure*}

\bsp	
\label{lastpage}
\end{document}